\documentclass[%
reprint,
superscriptaddress,
amsmath,amssymb,
aps,
pra,
]{revtex4-2}

\usepackage{graphicx}% Include figure files
\usepackage{dcolumn}% Align table columns on decimal point
\usepackage{bm}% bold math
\usepackage{color}
\usepackage{soul}
\begin{document}
    \title{Quantum multiparameter estimation enhanced by a topological phase transition}
	\author{Yu Yang}
	\email{yangyu1229@hotmail.com}
	\affiliation{Ministry of Education Key Laboratory for Nonequilibrium Synthesis and Modulation of Condensed Matter, Shaanxi Province Key Laboratory of Quantum Information and Quantum Optoelectronic Devices, School of Physics, Xi’an Jiaotong University, Xi’an 710049, China}
	\author{Haidong Yuan}
	\affiliation{Department of Mechanical and Automation Engineering, The Chinese University of Hong Kong, Shatin, Hong Kong SAR, China}
	\author{Fuli Li}
	\email{flli@xjtu.edu.cn}
	\affiliation{Ministry of Education Key Laboratory for Nonequilibrium Synthesis and Modulation of Condensed Matter, Shaanxi Province Key Laboratory of Quantum Information and Quantum Optoelectronic Devices, School of Physics, Xi’an Jiaotong University, Xi’an 710049, China}

\begin{abstract}
In quantum multiparameter estimation, multiple to-be-estimated parameters are encoded in a quantum dynamics system by a unitary evolution. 
As the parameters vary, the system may undergo a topological phase transition (TPT).
In this paper, we investigate two SU(2) TPT models and propose the singular behavior of the quantum metric tensor (QMT) around the TPT point as a tool for the simultaneous optimal estimation of multiple parameters.
We find that the proposed TPT sensing protocol  can achieve the same metrology performance as the quantum-control-enhanced one.
Moreover, the probe state of the TPT sensing protocol  is only the ground state of the Hamiltonian rather than the entangled state required in the control-enhanced one.
In addition, an adaptive multiparameter estimation strategy is developed for updating the estimated values till the desired quantum Cram$\acute{e}$r-Rao bound (QCRB) is approached.
Our work reinforces the connection between quantum multiparameter estimation and topology physics, with potential inspiration for quantum critical metrology.
\end{abstract}
\maketitle

\section{Introduction}
Recently, there has been a growing interest in studying quantum metrology from the geometrical perspective of quantum states~\cite{Changhao2022,Cai2022,Review_Goldman_2022,Provost1980,ReviewCarollo,Zeng2023}. 
The  motivation is to improve our comprehension about the physical nature behind quantum metrology and, more significantly,  to develop some novel approaches for enhancing the precision of parameter estimation.
These approaches are different from usual  metrology methods with employing the entangled probe state, designing the optimal measurement scheme, and optimizing the evolution of quantum dynamic systems. 
With the help of geometric quantities, it is possible to utilize the critical phenomena of quantum systems as a resource in quantum metrology, including the (high-order or nonlinear) exceptional point, (quantum or topological) phase transition, and so on.
This is precisely advocated in the so-called quantum critical metrology~\cite{critical,critical_review,critical_new}.

In quantum multiparameter estimation, multiple to-be-estimated parameters are encoded in a quantum dynamics system by a unitary evolution. 
The quantum dynamics system may experience a topological phase transition (TPT) as the parameters vary.
The TPT is identified by some topology invariants like the first Chern number and winding number.
Generally speaking, a physics system is extremely sensitive to the variations of parameters that drive the system to its phase transition point, thus the TPT can probably be used as a sensing tool to estimate the parameters.
In Ref.~\cite{Cai2022} some metrological bounds such as the quantum Cram$\acute{e}$r-Rao bound (QCRB) and the Holevo Cram$\acute{e}$r Rao bound (HCRB) across the TPT have been measured. 
The relation between the topology of Dirac Hamiltonian and quantum geometry has been presented in Ref.~\cite{Review_Goldman_2022}. 
The concept of the quantum volume of the Brillouin zone has been introduced in Ref.~\cite{Ozawa2021}, and it may be utilized to characterize the topological properties of the system.

Generators of translations in the parameter space are defined as Gauge potentials~\cite{Kolodrubetz2017,Demirplak2005}, 
which  covariance matrix is called the quantum geometric tensor (QGT) quantifying the distance between two neighboring quantum states over a quantum state manifold.
The real (symmetric) and imaginary (antisymmetric) components of the QGT are defined as the quantum metric tensor (QMT) and  Berry curvature, respectively.
The QGT, QMT, and Berry curvature have been measured in various experimental platforms, including the solid-state nitrogen-vacancy (NV) center in diamond~\cite{Cai2019}, superconducting circuits~\cite{Yu2019}, multiterminal Josephson junctions~\cite{Klees2020}, ultra-cold atoms~\cite{Flaschner2016}.
In quantum multiparameter estimation,  the estimation precision of multiple parameters is expressed by a covariance matrix that is bounded below by the well-known QCRB. 
The matrix-formed QCRB corresponds to the inverse of the quantum Fisher information matrix (QFIM), and each diagonal element of the QFIM is consistent with the quantum Fisher information (QFI) of the corresponding parameter. 
A major challenge in quantum multiparameter estimation is that  the estimation precisions of multiple parameters probably exist  trade-offs induced by the measurement incompatibility of the optimal protocols for the different parameters~\cite{Matsumoto2002,Review2019,Albarelli2020,Kok2021,Albarelli2022}. 
The presence of measurement incompatibility is caused by the Heisenberg uncertainty principle of quantum mechanics, which can be quantified by the self-defined figure of merit (FOM)~\cite{Belliardo2021,Chen2022}.
With the help of quantum geometrical notions one finds that the QMT (matrix) equals the 1/4-fold of the QFI (matrix),  and the FOM is associated with the QMT and  Berry curvature~\cite{Wang2016,Changhao2022,Review_Goldman_2022,Lu2021}.

In this paper, we present the geometrical properties of quantum states that are encoded in the sequentially-coding SU(2) dynamic system as shown in Fig.~\ref{Fig_scheme}(a), including the QGT, QMT, Berry curvature, and the first Chern number. 
The canonical model and the Su-Schrieffer-Heeger (SSH) model, two SU(2) TPT models are investigated in detail, and their topological features are identified by the first Chern number and the winding number, respectively.
We show that the QMT  displays a distinct peak in the vicinity of the TPT point, which can be used to develop a TPT sensing protocol.
Thus multiple parameters associated with the  TPT of the system can be simultaneously estimated with the individual highest estimation precision at the TPT point.
We discover that the proposed TPT sensing protocol can attain the same highest estimation precision as the quantum-control-enhanced protocol as shown in Fig.~\ref{Fig_scheme}(b)~\cite{Yuan2016,Hou2021,Hou2021PRL,Yang2022}.
Moreover, the probe state of our proposal  is only  the ground state of the Hamiltonian rather than the entangled state. 
In this way, the experimental burden of the probe state preparation can be relaxed.
Furthermore, an adaptive multiparameter estimation strategy is proposed and applied to the two SU(2) TPT models.
As shown in Fig.~\ref{Fig_scheme}(c) this strategy requires adaptive adjustment with updated estimated values till the attainable estimation precision approaches the desired QCRB.

The remainder of this paper is organized as follows.
Section~\ref{Sec:Geometry} gives an introduction to the geometry of parameterized quantum states and multiparameter estimation. 
In Sec.~\ref{Sec:topo}, we examine two SU(2) TPT models to demonstrate the benefits of TPT for multiparameter estimation.
The comparison of the TPT sensing protocol with control-enhanced sensing protocol is investigated in Sec.~\ref{Sec:VS}.
Sec.~\ref{Sec:scheme} presents an adaptive multiparameter estimation strategy based on the TPT for the two SU(2) TPT models.
Sec.~\ref{Sec:dissum} gives the summary for this work.\\

\section{Geometry of parameterized quantum state and multiparameter estimation}\label{Sec:Geometry}
For a collection of  unknown parameters 
$ \bm{\lambda}:=\{\lambda_1,\lambda_2,\cdots\lambda_{n}\} \in \mathcal{M}$ ($\mathcal{M}$ denotes the Hamiltonian parameters base manifold), a $\bm \lambda$-independent  initial probe state $|\psi\rangle$ acts on the $\bm \lambda$-dependent  unitary dynamics system $\hat{U}(\bm \lambda)$, the output state is $|\tilde{\psi}(\bm \lambda)\rangle=\hat{U}(\bm \lambda)|\psi\rangle$.
One can introduce the gauge potential as the generator of continuous unitary transformations, namely
\begin{eqnarray}\label{eq:GP}
i\hbar\partial_{\ell} |\tilde{\psi}(\bm \lambda)\rangle
={\mathcal{A}}_{\ell} |\tilde{\psi}(\bm \lambda)\rangle\;,
\end{eqnarray}
where $\partial_{\ell}$ means the derivative for the parameter $\lambda_\ell$ ($\ell\in \{1, 2, \cdots, n\}$), 
and the hermitian  gauge potential writes~\cite{Demirplak2005, Kolodrubetz2017}
\begin{eqnarray}\label{eq:def}
{\mathcal{A}}_{\ell}=i\hbar \partial_{\ell} \hat{U} (\bm \lambda) \hat{U}^\dagger(\bm \lambda)\;.
\end{eqnarray}
Eq.~(\ref{eq:def}) can be rewrriten as ${\mathcal{A}}_{\ell}=-i\hbar \hat{U}(\bm \lambda) \partial_{\ell} \hat{U}^\dagger(\bm \lambda)$ by employing $\partial_{\ell} (\hat{U}(\bm \lambda) \hat{U}^\dagger(\bm \lambda))=0$.
In the Heisenberg picture the gauge potential is 
\begin{eqnarray}\label{eq:Hdef}
\tilde{\mathcal{A}}_{\ell}=\hat{U}^\dagger(\bm \lambda) {\mathcal{A}}_{\ell} \hat{U}(\bm \lambda)
=i\hbar \hat{U}^\dagger(\bm \lambda) \partial_{\ell}\hat{U}(\bm \lambda)\;.
\end{eqnarray}
To simplify the following calculations, we set $\hbar=1$.

The (Abelian) QGT  based on the differential geometry describes the geometric characterizations of the wave function in the parameter space, which is defined as~\cite{Kolodrubetz2017,Yu2019,Note2022}
\begin{eqnarray}\label{QGT}
\chi_{\mu \nu}&=&\langle \partial_{\mu} \tilde{\psi}(\bm \lambda)| \partial_{\nu} \tilde{\psi}(\bm \lambda)\rangle \nonumber\\
&-&\langle \partial_{\mu} \tilde{\psi}(\bm \lambda)|\tilde{\psi}(\bm \lambda)\rangle \langle \tilde{\psi}(\bm \lambda)|\partial_{\nu} \tilde{\psi}(\bm \lambda)\rangle\;,
\end{eqnarray} 
for $\mu,\nu \in \{1,2,\cdots, n\}$.
Inserting Eq.~(\ref{eq:GP}) into Eq.~(\ref{QGT}), one has
\begin{eqnarray}\label{eq:AA}
\chi_{\mu \nu}&=&\langle \tilde{\psi}(\bm \lambda)|{\mathcal{A}}_{{\mu}} {\mathcal{A}}_{\nu}|\tilde{\psi}(\bm \lambda)\rangle\nonumber\\
&-&\langle \tilde{\psi}(\bm \lambda)|{\mathcal{A}}_{\mu}|\tilde{\psi}(\bm \lambda)\rangle
\langle \tilde{\psi}(\bm \lambda)|{\mathcal{A}}_{\nu}|\tilde{\psi}(\bm \lambda)\rangle\;.
\end{eqnarray}
By plugging Eq.~(\ref{eq:Hdef}) into Eq.~(\ref{eq:AA}), the counterpart of Eq.~(\ref{eq:AA}) in the Heisenberg picture reads 
\begin{eqnarray}\label{eq:AAnew}
\chi_{\mu \nu}=\langle {\psi}|{\tilde{\mathcal{A}}}_{{\mu}} \tilde{\mathcal{A}}_{\nu}|{\psi}\rangle-\langle {\psi}|\tilde{\mathcal{A}}_{\mu}|{\psi}\rangle
\langle {\psi}|\tilde{\mathcal{A}}_{\nu}|{\psi}\rangle\;.
\end{eqnarray}

The QMT (Fubini-Study metric tensor) over the parameter manifold is defined as the real part (or the symmetric part) of the QGT, i.e.
\begin{eqnarray}\label{eq:QMT}
g_{\mu \nu}=\text{Re}[\chi_{\mu \nu}]\;.
\end{eqnarray} 
Instituting Eq.~(\ref{eq:AAnew}) into Eq.~(\ref{eq:QMT}) one has
\begin{eqnarray}\label{eq:QM}
g_{\mu \nu}&=&\frac{\chi_{\mu \nu}+\chi_{\nu \mu}}{2}\nonumber\\
&=&\frac{1}{2}\langle \psi|\{\tilde{\mathcal{A}}_{\mu},  \tilde{\mathcal{A}}_{\nu}  \}|\psi\rangle \!-\! \langle \psi|\tilde{\mathcal{A}}_{\mu}|\psi\rangle \langle \psi|\tilde{\mathcal{A}}_{\nu}|\psi\rangle.
\end{eqnarray}
The imaginary part (or the anti-symmetric part) of the QGT is related to the Berry curvature as
\begin{eqnarray}\label{eq:BC}
\Omega_{\mu \nu}=-2 \text{Im}[\chi_{\mu \nu}]\;.
\end{eqnarray}
Inserting Eq.~(\ref{eq:AAnew}) into Eq.~(\ref{eq:BC}) we get
\begin{eqnarray}\label{eq:BCC}
\Omega_{\mu \nu}=i(\chi_{\mu \nu}-\chi_{\nu \mu})=i \langle \psi|[\tilde{\mathcal{A}}_{\mu},\tilde{\mathcal{A}}_{\nu}]|\psi\rangle\;.
\end{eqnarray}
The topological property of physics system can be characterized by the first Chern number~\cite{Kolodrubetz2017}
\begin{eqnarray}\label{eq:CC}
	{C}_{\mu \nu}=\frac{1}{2\pi} \int_\mathcal{S} \Omega_{\mu \nu}  d\lambda_{\mu} \wedge  d\lambda_\nu\;,
\end{eqnarray}
where $\wedge$ denotes the exterior (wedge) product, $\mathcal{S}$ represents the parameter space.

The QFIM with respect to the unknown parameters $\bm{\lambda}$ writes~\cite{Changhao2022,Cai2022,Review_Goldman_2022}
\begin{eqnarray}
{F}=4{G}\;,
\end{eqnarray}
where the $n \times n$ matrix ${G}$ is composed by the QMTs of Eq.~(\ref{eq:QM}).
The $\ell$-th diagonal element of ${F}$ corresponds to the QFI of the parameter $\lambda_\ell$.
One of the difficulties in quantum multiparameter estimation is that the highest estimation precisions for different parameters cannot be simultaneously reached in general.
This phenomenon of precision trade-offs  is referred to as the measurement incompatibility that results from the Heisenberg uncertainty relation of quantum mechanics.
The highest estimation precision is expressed by the matrix-formed QCRB: $ {F}^{-1}/M$, where $M$ is the number of times  the estimation procedure is repeated and ${F}^{-1}$ denotes the inverse matrix of the QFIM.
According to the Robertson-Schr$\ddot{o}$dinger uncertainty relation (see Appendix~\ref{App:incertainty_relation}), 
the FOM quantifying the measurement incompatibility can be depicted as~\cite{ Changhao2022,Lu2021,Carollo2019}
\begin{eqnarray}\label{eq:MI}
r_{\mu \nu}=\frac{\Omega_{\mu \nu}}{2\sqrt{\text{Det}\left[\mathcal{G}_{\mu \nu} \right]}}  \in [0,1]\;,
\end{eqnarray}
where $\text{Det}\left[\bullet\right]$ represents the matrix determinant, and 
\begin{eqnarray}\label{eq:g2}
	\mathcal{G}_{\mu \nu}=\left(\begin{matrix}
		g_{\mu \mu} & g_{\mu \nu}\\
		g_{\nu \mu} & g_{\nu \nu}
	\end{matrix}\right)\;.
\end{eqnarray}
is the $2 \times 2$ submatrix of ${G}$.
We remind that the FOM could have several distinct and useful definitions~\cite{Chen2022,Belliardo2021}.
Since the Berry curvature $\Omega_{\mu \nu}=0$ is equivalent to the weak communication condition~\cite{Carollo2019,Yang2022}, 
$r_{\mu \nu}=0$ means that parameters $\lambda_\mu$ and $\lambda_\nu$ can be simultaneously estimated.
However, $r_{\mu \nu}=1$ corresponds to the maximal estimation precision trade-off between $\lambda_\mu$ and $\lambda_\nu$.

\section{Quantum multiparameter estimation with TPT of SU(2) models}\label{Sec:topo}
\begin{figure}[!h]
	\centering
	\includegraphics[width=0.48\textwidth]{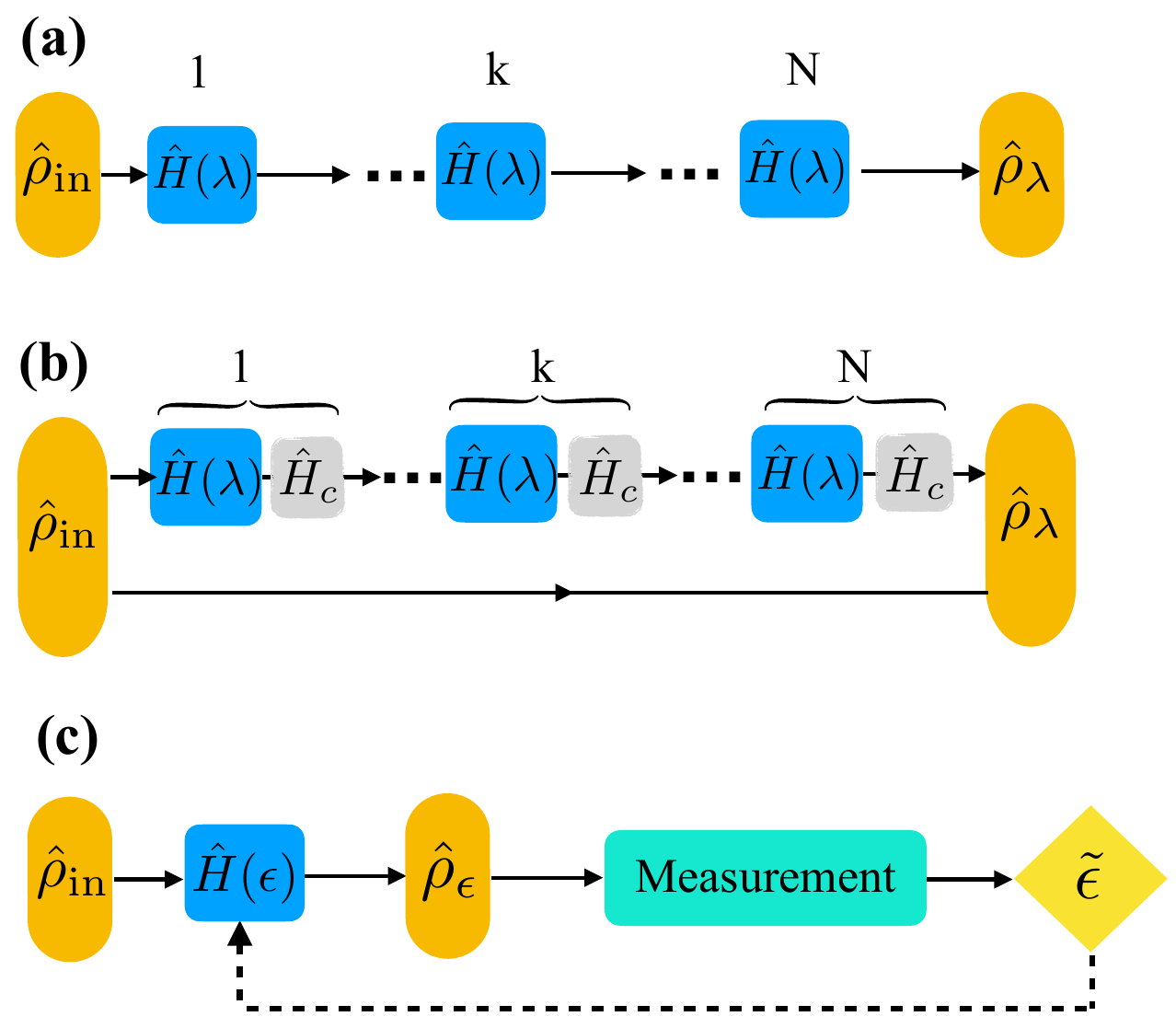}
	\caption
	{
	Panel (a): The N-order sequentially coding scheme is used for quantum multiparameter estimation.
	 The following measurement procedures and the data processing are not shown.
	$\hat{\rho}_{\text{in}}$ and $\hat{\rho}_{\bm \lambda}$ respectively denote the probe state and the encoded state for a set of to-be-estimated parameters ${\bm \lambda}=\{\lambda_1,\lambda_2, \lambda_3\}$.
	The whole dynamics evolution is divided into $N$ groups that each one includes an SU(2) parametrization process $\hat{H}({\bm \lambda})$.
	Panel (b): The control-enhanced sequentially coding scheme where every unitary cell includes not only  $\hat{H}({\bm \lambda})$ but also quantum control  $\hat{H}_c$. 
	An ancillary channel is added and has no interaction with the dynamics evolution. 
	Panel (c): An adaptive sensing scheme based on the topological phase transition (TPT) is used for quantum multiparameter estimation. 	
	The parameters associated with the TPT are denoted by $\bm{\epsilon}=\{\epsilon_i\}$ ($i\in[1,3]$), and the initial values of $\bm{\epsilon}$ are unknown and to be estimated. These TPT parameters  are usually  the subset of the parameters encoded in the Hamiltonian $\hat{H}(\bm{\lambda})$, i.e. $\bm{\epsilon} \in \bm{\lambda}$.  
	The encoded state $\hat{\rho}_{\bm \epsilon}$ is produced when the probe state $\hat{\rho}_\text{in}$ acts on the Hamiltonian $\hat{H}(\bm{\epsilon})$.	
	The TPT point  signals $\bm{\epsilon}$ being the critical values, with which the quantum metric tensor (i.e. the QFI) presents a peak. 	
	One continuously adjusts $\bm{\epsilon}$ from the initial points step by step until the TPT point is approached.
	The Hamiltonian $\hat{H}(\bm{\epsilon})$ is adaptively renewed with the  estimated values $\tilde{\bm{\epsilon}}$ after the measurement as the dashed line marked.
	According to the adjustment steps and the critical values, the initial values of $\bm{\epsilon}$ can be worked out.
}
	\label{Fig_scheme}
\end{figure}

In the $N$-order sequentially coding SU(2) unitary evolution depicted by Fig.~\ref{Fig_scheme}(a), 
the whole unitary transformation from $\hat{\rho}_\text{in}$ to $\hat{\rho}_\lambda$ is 
\begin{eqnarray}
\hat{U}=\left(e^{-it\hat{H}(\bm \lambda)}\right)^N=e^{-iH(\bm \lambda)T}\;,
\end{eqnarray}
where $T=t N$ is the total evolution time with an integer $N$. The
 generic time-independent Hamiltonian is
\begin{eqnarray}\label{eq:su(2)}
	\hat{H}(\bm \lambda)=\mathbf{X} \cdot \vec{J}\;,
\end{eqnarray}
where $\mathbf{X}=(X_1(\bm \lambda),X_2(\bm \lambda),X_3(\bm \lambda))$ is a three-dimensional vector, $X_l(\bm\lambda)$ is a function of $\bm\lambda$ with $l=1,2,3$, $\vec{J}=(\hat{j}_1,\hat{j}_2,\hat{j}_3)$ are three generators of SU(2) algebra obeying the commutation relation $\left[ \hat{j}_m,\hat{j}_n\right]=i\xi_{mkl} \hat{j}_l$ with the Levi-Civita symbol $\xi_{mkl}$.
In this SU(2) parameterization process, the initial probe state is assumed to be a single-qubit pure state $\hat{\rho}_\text{in}=\hat{I}/2 +\vec{r}_\text{in} \cdot \vec{J}$ with the Bloch vector $\vec{r}_\text{in}$ ($||\vec{r}_\text{in}||=1$), $\hat{I}$ is an identity operator.
The QMT associated with parameters $\lambda_\mu$, $\lambda_{\nu}$ can be expressed by (see Appendix~\ref{Subsec:without})
\begin{eqnarray}\label{eq:gnew}
	g_{\mu \nu}=\frac{|\mathbf{Y}_\mu||\mathbf{Y}_\nu|}{4} [(\vec{e}_\mu \cdot \vec{e}_{\nu})-(\vec{e}_\mu \cdot \vec{r}_\text{in})(\vec{e}_{\nu} \cdot \vec{r}_\text{in})]\;,
\end{eqnarray}
with the unit vectors
\begin{eqnarray}\label{eq:vector}
&&\vec{e}_\ell=\frac{1}{|\mathbf{Y}_\ell|} \left\lbrace 
-T ( \partial_\ell \mathbf{X})+\frac{|\partial_\ell \mathbf{X}||\sin \alpha_\ell|}{|\mathbf{X}|}\right.\nonumber\\
&&\times\left.\Big\{[ \sin(T|\mathbf{X}|)-T|\mathbf{X}|] \vec{v}_{\ell,2} +\left[1-\cos(T|\mathbf{X}|) \right]  \vec{v}_{\ell,1}
\Big\} \right\rbrace, \nonumber\\\\
&&\vec{v}_{\ell,1}=\frac{\mathbf{X} \times \partial_\ell \mathbf{X}}{\left| \mathbf{X}\right| \left| \partial_\ell \mathbf{X}\right| \sin \alpha_\ell }\;,
\vec{v}_{\ell,2}=\frac{\mathbf{X} \times\left(\mathbf{X} \times \partial_\ell \mathbf{X}\right)}{\left| \mathbf{X}\right|^2 \left| \partial_\ell \mathbf{X}\right| | \sin \alpha_\ell|},
\end{eqnarray}
where $\alpha_\ell$  is the angle between vectors $\mathbf{X}$ and $\partial_\ell \mathbf{X}  \; (\partial_\ell \mathbf{X}:=\partial \mathbf{X}/\partial \lambda_\ell)$,  
and
\begin{eqnarray}\label{eq:Y}
	|\mathbf{Y}_\ell|\!=\!\sqrt{T^2 |\partial_\ell \mathbf{X} |^2 \cos^2\alpha_\ell\!+\!\frac{4|\partial_\ell \mathbf{X}|^2 \sin^2 \alpha_\ell}{|\mathbf{X}|^2} \sin^2 \left(\frac{T|\mathbf{X}|}{2} \right)}, \nonumber\\
\end{eqnarray}
with $\ell \in \{\mu,\nu\}$.
For $\lambda_\mu=\lambda_\nu$, Eq.~(\ref{eq:gnew}) can be simplified as
\begin{eqnarray}\label{eq:qmt}
	g_{\mu \mu}=\frac{|\mathbf{Y}_\mu|^2}{4} [1-(\vec{e}_\mu \cdot \vec{r}_\text{in})^2]\;.
\end{eqnarray}
The corresponding Berry curvature and the first Chern number are worked out as
\begin{eqnarray}\label{eq:bc}
	\Omega_{\mu \nu}&=&-\frac{|\mathbf{Y}_\mu| |\mathbf{Y}_\nu|}{2} (\vec{e}_\mu \times \vec{e}_\nu) \cdot \vec{r}_\text{in}\;,\\
	C_{\mu \nu}&=&
	\frac{-1}{4\pi} \int_{S^2} |\mathbf{Y}_\mu| |\mathbf{Y}_{\nu}| (\vec{e}_\mu \times \vec{e}_{\nu}) \cdot \vec{r}_\text{in} d\lambda_\mu \wedge d\lambda_{\nu},
\end{eqnarray}
where $S^2$ denotes the Bloch sphere.

The SU(2) coding dynamics system (\ref{eq:su(2)}) may experience a topological phase transition as the parameters vary.
In general, a physics system is extremely sensitive to the variations of some parameters around its phase transition point. 
Thus the phase transition may provide us with a useful tool for quantum sensing.
To deeply investigate this possibility, in the following section we take two typical SU(2) TPT models as examples to show that parameters related to the TPT can be simultaneously estimated with the individual highest precision at the TPT point.

\subsection{The canonical model with TPT characterized by the first Chern number}\label{Subsec:model1}
We consider one canonical model 
\begin{eqnarray}\label{eq:TH}
\hat{H}=\vec{m} \cdot \vec{J}\;,
\end{eqnarray}
with
\begin{eqnarray}\label{eq:m}
	\vec{m}=2H_0(\sin\theta \cos\phi, \sin \theta \sin\phi, \cos\theta+r)\;,
\end{eqnarray}
where $\theta \in [0,\pi]$, $\phi \in [0,2\pi]$, $r$ is a tunable parameter.
The conditions $|r|<1$ and $|r|>1$ correspond to topologically non-trivial and trivial regimes,
$r=1$ and $\theta=\pi$ give the singular behaviors in the first Chern number and Berry curvature, i.e. the TPT occurs~\cite{Cai2019, Cai2022}. 
After the Jordan-Wigner and Fourier transformations to a two-dimensional momentum space, this model may be used to represent a many-body XY spin chain~\cite{momentum}.
Particularly when $r=0$, the Hamiltonian (\ref{eq:TH}) is typically used to metrology the amplitude and direction of an unknown magnetic field~\cite{Hou2021,Pang2014}.

The Hamiltonian (\ref{eq:TH}) is loaded by the multiparameter estimation scheme as shown in Fig.~\ref{Fig_scheme} (a).
The initial probe state is a single-qubit pure state with the Bloch vector $\vec{r}_\text{in}$ ($||\vec{r}_\text{in}||=1$). 
If  $\vec{e}_\theta \cdot \vec{r}_\text{in}=\vec{e}_\phi \cdot \vec{r}_\text{in}=\vec{e}_r \cdot \vec{r}_\text{in}=0$ can be satisfied, 
according to Eq.~(\ref{eq:qmt}) we can obtain the maximal QMTs of  $\theta,\phi,r$ as 
\begin{widetext}
\begin{eqnarray}
g^{(M)}_{\theta \theta}&=& \frac{ r^2 T^2 \sin^2\theta}{1+r^2+2r \cos\theta} +\left(\frac{1+r \cos\theta}{1+r^2+2r \cos\theta}\right)^2 \sin^2[T \sqrt{1+r^2+2r \cos\theta}] \;,\label{eq:thetanew}\\
g^{(M)}_{\phi \phi}&=&\left(\frac{\sin^2 \theta }{1+r^2+2r\cos\theta}\right) \sin^2[ T \sqrt{1+r^2+2r \cos\theta}]\;,\label{eq:phinew}\\
g^{(M)}_{r r}&=&\frac{T^2(r+\cos\theta)^2 }{1+r^2+2r\cos\theta}+\left(\frac{\sin\theta}{1+r^2+2r \cos\theta}\right)^2 \sin^2[T\sqrt{1+r^2+2r\cos\theta}]\label{eq:gr}\;,
\end{eqnarray}
where $H_0=1$ is set for the simplification.
In Figs.~\ref{Figc}(a)-(c), Eqs.~(\ref{eq:thetanew})-(\ref{eq:gr}) are plotted with the yellow surfaces for the given value of $\theta=\pi$.
Fig.~\ref{Figc}(d) further displays the relation among $g_{\theta \theta}^{(M)}$, $\theta$ and $r$.
Figs.~\ref{Figc}(e)-(f) separately give the functional dependence of $g_{\theta \theta}^{(M)}$ upon $\theta$ and $r$ for the given values of $r=\{1,1-0.1,1-0.5\}$ and $\theta=\{\pi,\pi-0.1,\pi-0.5\}$.
In Figs.~\ref{Figc}(a) and (d), one especially notices that $g_{\theta\theta}^{(M)}$ exhibits a peak when $\theta=\pi$ and $r=1$, i.e. the estimation precision of $\theta$ reaches to its maximum at the TPT point.
\begin{figure}[!h]
\centering
\includegraphics[width=0.75\textwidth]{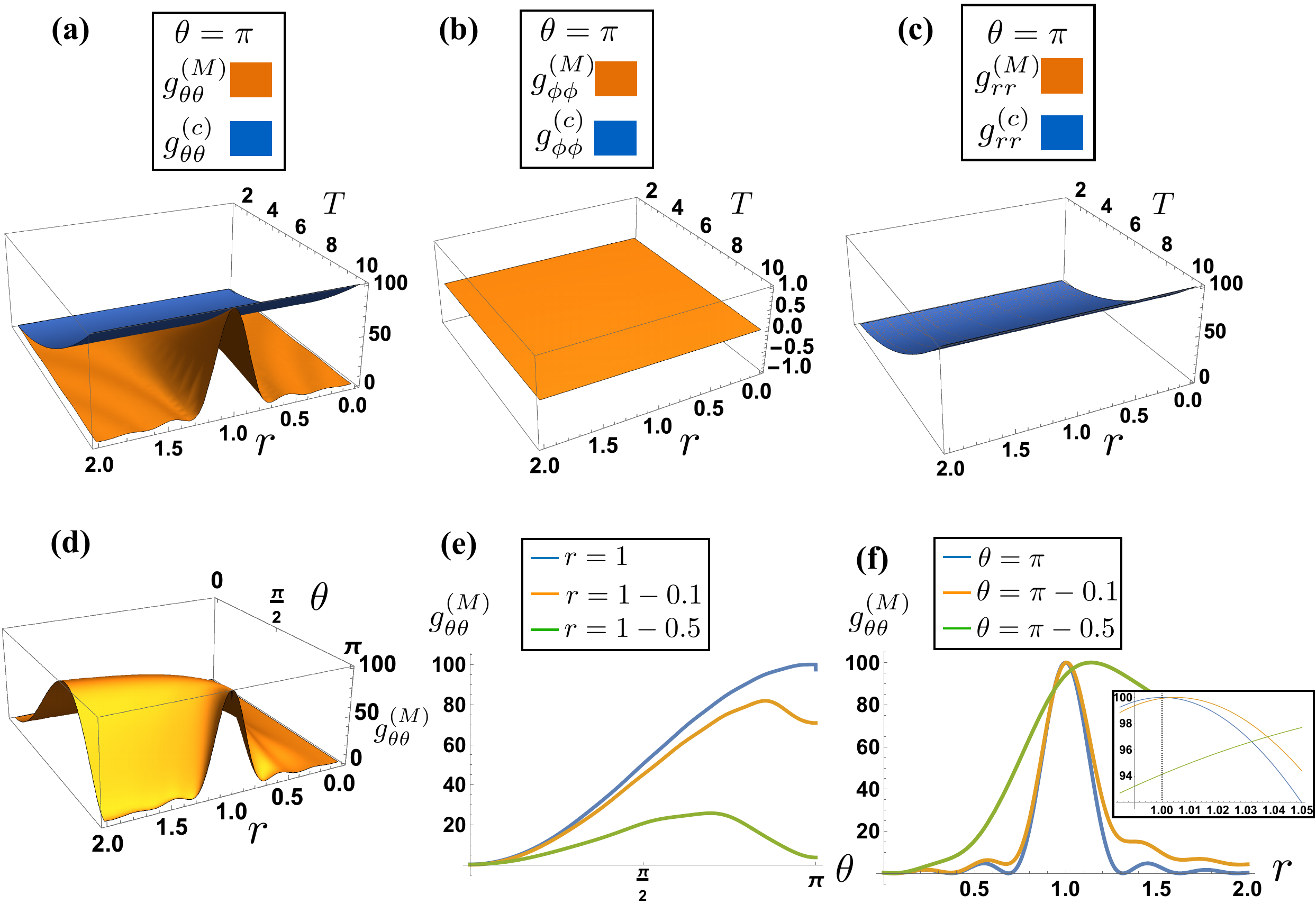}
\caption
{Panels (a)-(c): The relation among the maximal QMTs $g_{\theta \theta}^{(M)}$, $g_{\phi \phi}^{(M)}$, $g_{rr}^{(M)}$ of Eqs.~(\ref{eq:thetanew})-(\ref{eq:gr}) ($g_{\theta \theta}^{(c)}$, $g_{\phi \phi}^{(c)}$, $g_{rr}^{(c)}$ of Eqs.~(\ref{eq:QMTtheta})-(\ref{eq:QMTr})), the total evolution time $T$ and the parameter $r$ are plotted with the yellow (blue) surfaces for the given value of $\theta=\pi$, respectively. 
In panels (b) and (c), two surfaces both overlap completly and have no dependence on the topological parameter $r$. 
Panel (d) displays the relation among $g_{\theta \theta}^{(M)}$, $\theta$ and $r$. 
The functional dependence of $g_{\theta \theta}^{(M)}$ upon $\theta$ and $r$ for the given values of $r=\{1,1-0.1,1-0.5\}$ and $\theta=\{\pi,\pi-0.1,\pi-0.5\}$ are individually plotted in panels~(e)-(f).
The inset figure is also exhibited in panel (f) to enlarge the results $g_{\theta \theta}^{(M)}=\{100,99.9271,94.1155\}$  with the given values of $\theta=\{\pi,\pi-0.1,\pi-0.5\}$ for $r=1$.
The highest estimation precision of $\theta$ depends on the TPT condition, i.e. $\theta=\pi,r=1$, which is exhibited by a QMT peak around the TPT point in panel (a).
Here  $T=10$ is set for the simulation.}
\label{Figc}
\end{figure}

As a more specific example, we take the ground state of the Hamiltonian (\ref{eq:TH}) as the initial probe state.
The corresponding Bloch vector is $\vec{r}^\prime_\text{in}=(\sin\theta' \cos\phi,\sin\theta' \sin\phi,\cos\theta')$ with $\theta'=\arccos\left[{(\cos\theta +r)/}{\sqrt{1+r^2+2r \cos\theta}}\right]$.
Plugging Eq.~(\ref{eq:m}) into Eqs.~(\ref{eq:vector})-(\ref{eq:Y}), we have
\begin{eqnarray}
\vec{e}_\theta \cdot \vec{r}^\prime_\text{in}&=&\frac{rT\sqrt{1+r^2+2r \cos\theta}\sin\theta}{\sqrt{r^2T^2(1+r^2+2r \cos\theta)\sin^2 \theta}+(1+r \cos\theta)^2\sin^2[T\sqrt{1+r^2+2r \cos\theta}]}\;,\\
\vec{e}_\phi \cdot \vec{r}^\prime_\text{in}&=&0\;,\\
\vec{e}_r \cdot \vec{r}^\prime_\text{in}&=&\frac{-T(r+\cos\theta)\sqrt{1+r^2+2r \cos \theta}}{\sqrt{T^2(r+\cos\theta)^2(1+r^2+2r \cos\theta)+\sin^2 \theta \sin^2[T\sqrt{1+r^2+2r\cos\theta}]}}\;.
\end{eqnarray}
Inserting these results into Eqs.~(\ref{eq:gnew}) and (\ref{eq:qmt}), we obtain the QMT matrix
\begin{eqnarray}
{G}=\left(\begin{matrix}\label{eq:Gmatrix}
\frac{(1+r \cos\theta)^2 \sin^2[T\sqrt{1+r^2+2r \cos\theta}]}{(1+r^2+2r \cos\theta)^2} &0&\frac{-(1+r \cos\theta)\sin\theta \sin^2[T\sqrt{1+r^2+2r \cos\theta}]}{(1+r^2+2r \cos\theta)^2}\\
0 & \frac{\sin^2\theta \sin^2[T\sqrt{1+r^2+2r \cos\theta}]}{1+r^2+2r \cos\theta} & 0\\
\frac{(1+r \cos\theta)\sin\theta \sin^2[T\sqrt{1+r^2+2r \cos\theta}]}{(1+r^2+2r \cos\theta)^2} & 0& \frac{\sin^2 \theta \sin^2 [T \sqrt{1+r^2+2r \cos\theta}]}{(1+r^2+2r \cos\theta)^2}
\end{matrix}\right)\;.
\end{eqnarray}
In the limit of $\theta \to \pi$ and $r \to 1$, it reduces to
\begin{eqnarray}\label{eq:LimitG}
\lim\limits_{\substack{\theta\to\pi \\ r\to1}}{G}=\left(\begin{matrix}
T^2 & 0 &0\\
0 & 0& 0\\
0 & 0 &T^2
\end{matrix}\right)\;.
\end{eqnarray}
Eq.~(\ref{eq:LimitG}) indicates that parameters $\theta$ and $r$ can be estimated with the individual highest estimation precision (i.e. the Heisenberg scaling $1/T$) in the vicinity of TPT.
Since the condition of generating the TPT does not refer to the parameter $\phi$, naturally, we cannot extract any information about $\phi$ by virtue of the TPT as shown in Eq.~(\ref{eq:LimitG}) and Fig.~\ref{Figc}(b).
With some algebraic operations, Eq.~(\ref{eq:bc}) gives the  Berry curvature matrix as
\begin{eqnarray}\label{eq:BCmatrix}
{\Omega}=\left(\begin{matrix}
0 & \frac{2(1+r \cos\theta)\sin\theta \sin^2[T\sqrt{1+r^2+2r \cos\theta}]}{(1+r^2+2r \cos\theta)^{3/2}} & 0\\
- \frac{2(1+r \cos\theta)\sin\theta \sin^2[T\sqrt{1+r^2+2r \cos\theta}]}{(1+r^2+2r \cos\theta)^{3/2}} & 0 & \frac{2\sin^2\theta \sin^2[T\sqrt{1+r^2+2r \cos\theta}]}{(1+r^2+2r \cos\theta)^{3/2}}\\
0 & -\frac{2\sin^2\theta \sin^2[T\sqrt{1+r^2+2r \cos\theta}]}{(1+r^2+2r \cos\theta)^{3/2}} & 0
\end{matrix}\right) \;,
\end{eqnarray}
where $\Omega_{\theta r}=\Omega_{r \theta}=0$ since $(\vec{e}_\theta \times \vec{e}_r)\cdot \vec{r}^\prime_\text{in}=0$.
The implied physical meaning of $\Omega_{\theta r}=0$ is that parameters $\theta$ and $r$ can be estimated simultaneously, but the individual estimation precision is not the highest without the help of TPT. 
The matrix element $\Omega_{\theta \phi}$ of Eq.~(\ref{eq:BCmatrix}) is plotted in Fig.~\ref{FigBC}(a) with the blue curve and a series of oscillations over the time $T$ are displayed.
Inserting the expression of $\Omega_{\theta \phi}$ into Eq.~(\ref{eq:CC}), we get the first Chern number ${C}_{\theta \phi}$ that is plotted in Fig.~\ref{FigBC}(b) with the blue curve and the similar oscillations over the time $T$ are accompanied.
To investigate the characterizations of Berry curvature, we use a mathematical processing method called ``coarse graining"~\cite{coarse,coarsenew}, which averages the Berry curvature over $T$ as $ \bar{\Omega}:=\frac{1}{T}\int_{t-T/2}^{t+T/2} dt\; \Omega$. Thus Eq.~(\ref{eq:BCmatrix}) is renewed as
\begin{eqnarray}\label{eq:ave_BCmatrix}
\bar{{\Omega}}=\left(\begin{matrix}
0 & \frac{(1+r \cos\theta)\sin\theta}{(1+r^2+2r \cos\theta)^{3/2}} &0 \\
-\frac{(1+r \cos\theta)\sin\theta}{(1+r^2+2r \cos\theta)^{3/2}} & 0 & \frac{\sin^2\theta}{(1+r^2+2r \cos\theta)^{3/2}} \\
0 & -\frac{\sin^2\theta}{(1+r^2+2r \cos\theta)^{3/2}} & 0
\end{matrix}\right)\;.
\end{eqnarray}
\end{widetext}
In the limit of $\theta \to \pi$ and $r \to 1$, both $\Omega$ and $\bar{\Omega}$ reduce to
\begin{eqnarray}\label{eq:newBC}
	\lim\limits_{\substack{\theta\to\pi \\ r\to1}}	{\Omega}=	\lim\limits_{\substack{\theta\to\pi \\ r\to1}}	\bar{\Omega}=\mathbf{0}\;,
\end{eqnarray}
where $\mathbf{0}$ represents a zero matrix.
The matrix element $\bar{\Omega}_{\theta \phi}$ of Eq.~(\ref{eq:ave_BCmatrix}) is also plotted in Fig.~\ref{FigBC}(a) with the yellow curve.
Substituting the form of $\bar{\Omega}_{\theta \phi}$ into Eq.~(\ref{eq:CC}), we get the coarse graining counterpart of the first Chern number
\begin{eqnarray}\label{eq:aveChern}
\bar{{C}}_{\theta \phi}=\frac{[\text{sgn}(r-1)-1] \text{sgn}(r^2-1)}{\text{sgn}(r-1)}\;,
\end{eqnarray}
where \text{sgn}($\bullet$) represents a sign function.  Eq.~(\ref{eq:aveChern}) is plotted in Fig.~\ref{FigBC}(b) with the yellow curve.
The TPT point is represented by $r=1$ and $\theta=\pi$, as shown in Figs.~\ref{FigBC}(a)-(b).
To clarify the relation between our results  and the existing results~\cite{Cai2022,Cai2019,Goldman2018}, some discussions are presented  in Appendix~\ref{App:comparison}. 
\begin{figure}[!h]
	\centering
	\includegraphics[width=0.35\textwidth]{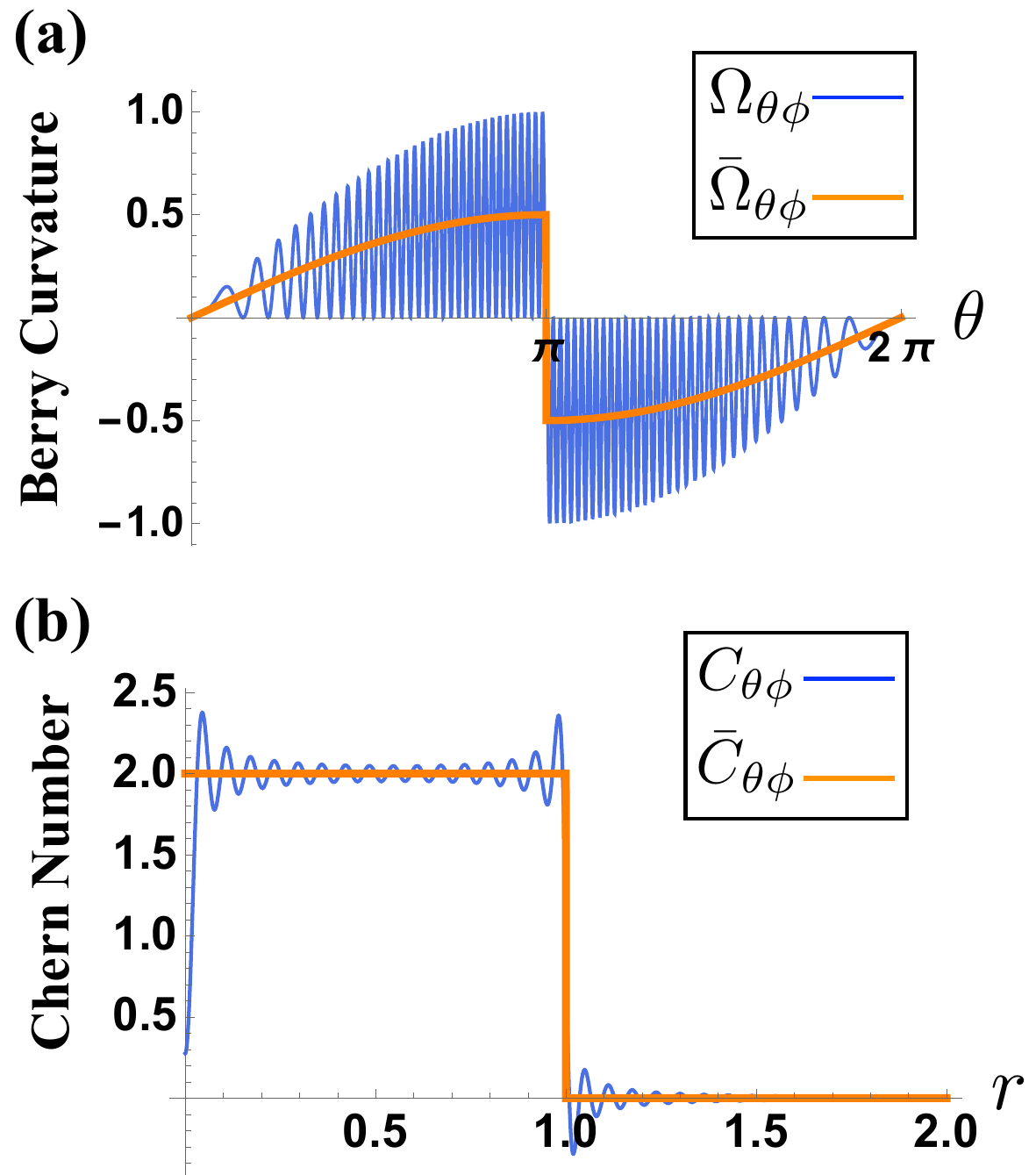}
	\caption{
The functional dependence of the Berry curvature $\Omega_{\theta \phi}$  of Eq.~(\ref{eq:BCmatrix}) upon the parameter $\theta$ is plotted in the panel (a) with the blue curve. 
Panel (b) shows the functional dependence of the corresponding first Chern number $C_{\theta \phi}$  upon the parameter $r$ with the blue curve. 
After averaging the Berry curvature over $T$, the functional dependence of the Berry curvature $\bar{\Omega}_{\theta \phi}$  of Eq.~(\ref{eq:ave_BCmatrix}) upon the parameter $\theta$ is also plotted in the panel (a) with the yellow curve.
The coarse-grained first Chern number $\bar{C}_{\theta \phi}$ (\ref{eq:aveChern}) is plotted in the panel (b) with the yellow curve as well.
Here $T=50$ is set for the simulation.}
\label{FigBC}
\end{figure}
Substituting Eqs.~(\ref{eq:Gmatrix}) and (\ref{eq:BCmatrix}) into (\ref{eq:MI}), one has the FOM matrix
\begin{eqnarray}\label{eq:rmatrix}
R=\left(\begin{matrix}
0& 1 & 0\\
-1 & 0 & 1\\
0 & -1 & 0
\end{matrix}\right)\;.
\end{eqnarray}
Eq.~(\ref{eq:rmatrix}) implies that the parameters $\{\theta, r\}$ can be simultaneously estimated, and this has no dependence on the TPT. The optimal measurement scheme for $\{\theta, r\}$ is also presented in Appendix~\ref{App:meas}.

\subsection{The SSH model with TPT characterized by winding number}\label{Subsec:model2}
The Su-Schrieffer-Heeger model is generally used to study the case of fermions (like spin-polarized electrons) hopping on a one-dimensional lattice chain where multiple unit cells are  sequentially organized.
The  chiral symmetry  bulk Hamiltonian writes~\cite{topological_book}
\begin{eqnarray}\label{eq:SSH}
	\hat{H}^\prime=\vec{\nu} \cdot \vec{J}\;,
\end{eqnarray}
with
\begin{eqnarray}\label{eq:vv}
\vec{\nu}=2(v+w\cos k, w \sin k,0)\;,
\end{eqnarray}
where $v$, $w$ represent the intracell and intercell hopping amplitudes for $k \in [-\pi,\pi]$, respectively.
The TPT takes place when $v=w$ and $k=\pm \pi$, which produce the singular behavior in the winding number and the gap-closing point of energy bands~\cite{topological_book}.

The Hamiltonian (\ref{eq:SSH}) is loaded by the multiparameter estimation scenario as shown in Fig.~\ref{Fig_scheme} (a). 
The initial probe state is still a single-qubit pure state but with the different Bloch vector $\tilde{r}_\text{in}$ ($||\tilde{r}_\text{in}||=1$).
If $\vec{e}_v \cdot \tilde{r}_\text{in}=\vec{e}_w \cdot \tilde{r}_\text{in}=\vec{e}_k \cdot \tilde{r}_\text{in}=0$ can be satisified, 
according to Eq.~(\ref{eq:qmt}) we can get the maximal QMTs of  $v,w,k$ as 
\begin{widetext}
\begin{eqnarray}
g^{(M)}_{v v}&=&\frac{T^2 (v+w \cos k)^2}{v^2+w^2+2 v w \cos k}+\left(\frac{w \sin k}{v^2+w^2+2 v w \cos k}\right)^2 \sin^2 [T\sqrt{v^2+w^2+2 v w \cos k}],\label{eq:ncv}\\
g^{(M)}_{w w}&=&\frac{T^2 (w+v \cos k)^2}{v^2+w^2+2vw \cos k}+\left(\frac{v \sin k}{v^2+w^2+2 v w \cos k}\right)^2 \sin^2 [T \sqrt{v^2+w^2+2v w \cos k}],\label{eq:ncw}\\
g^{(M)}_{k k}&=&\frac{T^2 w^2 v^2 \sin^2 k}{v^2+w^2+2v w \cos k}+\left(\frac{w+v \cos k}{v^2+w^2+2 v w \cos k}\right)^2 \sin^2[T\sqrt{v^2+w^2+2v w \cos k}].\label{eq:nck}
\end{eqnarray}
In Figs.~\ref{FigSSH}(a)-(c), Eqs.~(\ref{eq:ncv})-(\ref{eq:nck}) are plotted with the yellow surfaces for the given value of $k=\pi$.
Fig.~\ref{FigSSH}(d) further displays the relation among $g_{k k}^{(M)}$, $k$ and $v$ for $w=1$.
Figs.~\ref{FigSSH}(e)-(f) separately give the functional dependence of $g_{kk}^{(M)}$ upon $k$ and $v$ for the given values of $v=\{1,1-0.1,1-0.5\}$ and $k=\{\pi,\pi-0.1,\pi-0.5\}$, and $w=1$ is set.
Especially Figs.~\ref{FigSSH}(c) and (d) shows that $g_{kk}^{(M)}$ exhibits a peak when $k=\pi$ and $v=w$, i.e. the estimation precision of $k$ reaches to its maximum at the TPT point.
\begin{figure}[!h]
\centering
\includegraphics[width=0.75\textwidth]{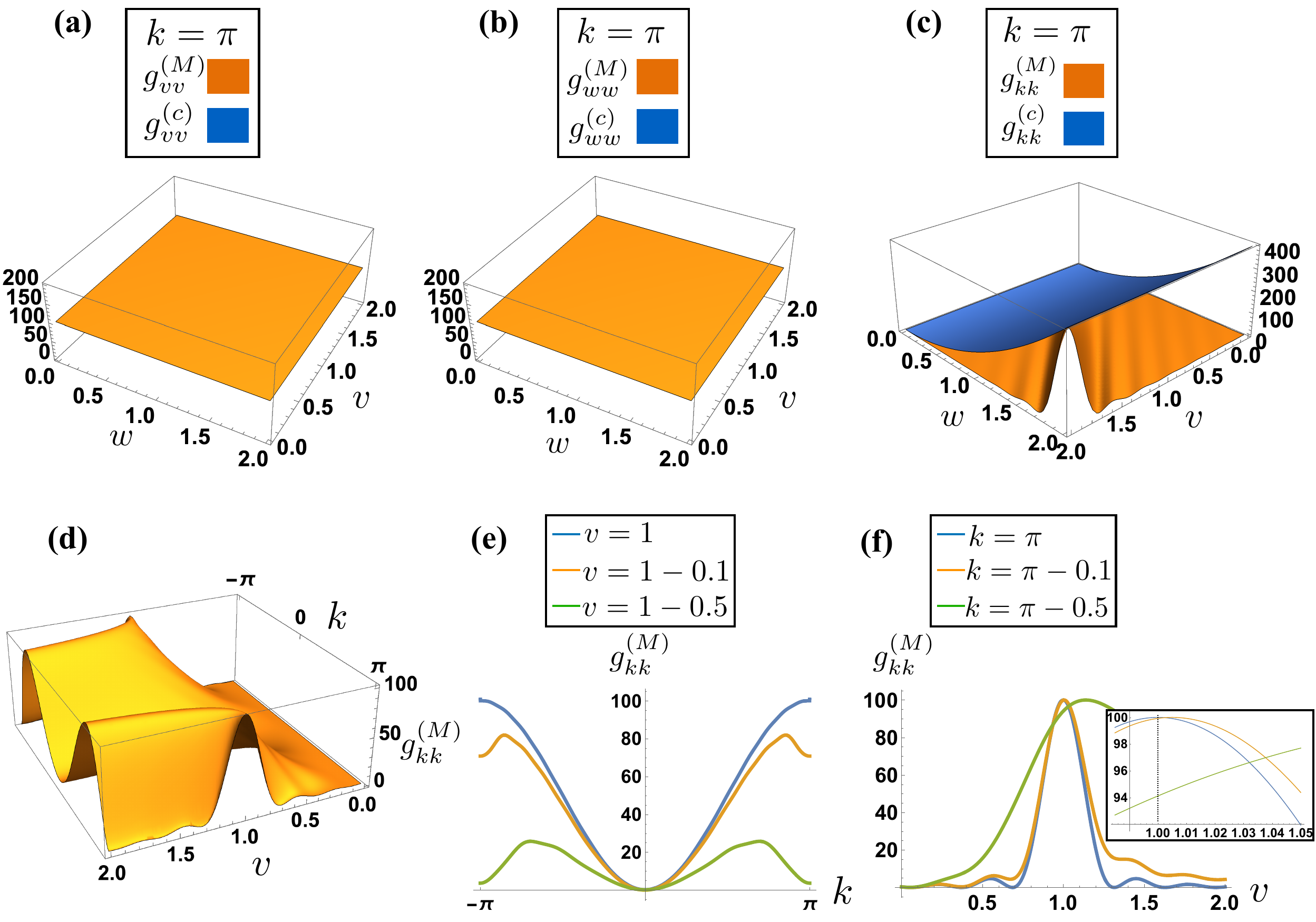}
\caption
{Panels (a)-(c): The relation among the maximal QMTs $g_{v v}^{(M)}$, $g_{w w}^{(M)}$, $g_{kk}^{(M)}$ of  Eqs.~(\ref{eq:ncv})-(\ref{eq:nck}) ($g_{v v}^{(c)}$, $g_{w w}^{(c)}$, $g_{kk}^{(c)}$ of Eqs.~(\ref{eq:cv})-(\ref{eq:ck})), the parameters $v$ and $w$ are plotted with the yellow (blue) surfaces for the given value of $k=\pi$, respectively. 
In panels (a) and (b), two surfaces both overlap completly and have no dependence on the relation of parameters $v,w$. 
Panel (d) displays the relation among $g_{kk}^{(M)}$, $k$ and $v$ for $w=1$. 
Panels (e)-(f) separately give the functional dependence of $g_{kk}^{(M)}$ upon $k$ and $v$ for the given values of $v=\{1,1-0.1,1-0.5\}$ and $k=\{\pi,\pi-0.1,\pi-0.5\}$, and $w=1$ is set.
The inset figure is also displayed in panel (f) to enlarge the results $g_{kk}^{(M)}=\{100,99.9271,94.1155\}$  with the given values of $k=\{\pi,\pi-0.1,\pi-0.5\}$ for $v=w=1$.
The highest estimation precision of $k$ depends on the TPT condition, i.e. $k=\pi,v=w$, which is shown as a QMT peak around the TPT point in panel (c).
Here $T=10$ is set for the simulation.}
\label{FigSSH}
\end{figure}

Here we take the ground state of the Hamiltonian (\ref{eq:SSH}) as the initial probe state with the Bloch vector  $\tilde{r}_\text{in}^\prime=1/\sqrt{v^2+w^2+2v w \cos k}(v+w\cos k, w\sin k,0)$.
Inserting Eq.~(\ref{eq:vv}) into Eqs.~(\ref{eq:vector})-(\ref{eq:Y}), we obtain
\begin{eqnarray}
\vec{e}_v \cdot \tilde{r}^\prime_\text{in}&=&\frac{-T(v+w \cos k)}{\sqrt{T^2(v+w \cos k)^2+\frac{w^2 \sin^2 k \sin^2[T \sqrt{v^2+w^2+2v w \cos k}]}{v^2+w^2+2 v w \cos k}}},\\
\vec{e}_w \cdot \tilde{r}^\prime_\text{in}&=&\frac{-T(w+v \cos k)}{\sqrt{T^2(w+v \cos k)^2+\frac{v^2 \sin^2 k \sin^2[T\sqrt{v^2+w^2+2vw \cos k}]}{v^2+w^2+2vw \cos k}}},\\ 
\vec{e}_k \cdot \tilde{r}^\prime_\text{in}&=&\frac{Tv\sin k}{\sqrt{T^2 v^2 \sin^2 k+\frac{(w+v \cos k)^2 \sin^2[T \sqrt{v^2+w^2+2vw \cos k}]}{v^2+w^2+2vw\cos k}}}. 
\end{eqnarray}
By inserting these results into Eqs.~(\ref{eq:gnew}) and (\ref{eq:qmt}), the QMT matrix writes
\begin{eqnarray}\label{eq:Gprime}
\hspace{-10mm}
	{G}'\!=\!
	\!\left(\!\begin{matrix}
		\frac{w^2\sin^2k \sin^2(T\sqrt{v^2+w^2+2v w \cos k})}{(v^2+w^2+2v w \cos k)^2} & \frac{-vw\sin^2k\sin^2(T\sqrt{v^2+w^2+2v w \cos k})}{(v^2+w^2+2v w \cos k)^2} & \frac{-w^2 (w+v \cos k)\sin k \sin^2(T\sqrt{v^2+w^2+2v w \cos k})}{(v^2+w^2+2v w \cos k)^2}\\\\
		\frac{-vw\sin^2k\sin^2(T\sqrt{v^2+w^2+2v w \cos k})}{(v^2+w^2+2v w \cos k)^2} & \frac{v^2 \sin^2k\sin^2(T\sqrt{v^2+w^2+2v w \cos k})}{(v^2+w^2+2v w \cos k)^2} & \frac{vw(w+v \cos k)\sin k \sin^2(T\sqrt{v^2+w^2+2v w \cos k})}{(v^2+w^2+2 v w \cos k)^2}\\\\
		\frac{-w^2 (w+v \cos k)\sin k \sin^2(T\sqrt{v^2+w^2+2v w \cos k})}{(v^2+w^2+2v w \cos k)^2} &
		\frac{vw(w+v \cos k)\sin k \sin^2(T\sqrt{v^2+w^2+2v w \cos k})}{(v^2+w^2+2 v w \cos k)^2} & \frac{w^2(w+v \cos k)^2 \sin^2 (T\sqrt{v^2+w^2+2vw\cos k})}{(v^2+w^2+2 vw\cos k)^2}
	\end{matrix}\!\right)\!.\! \nonumber\\
\end{eqnarray}
\end{widetext}
In the limit of $k \to \pi$ and $v \to w$, it reduces to
\begin{eqnarray}\label{eq:newG}
\lim\limits_{\substack{k\to \pi \\ v \to w}}{G}'=\left(\begin{matrix}
T^2 &0 & 0\\
0 & T^2 & 0\\
0 & 0& T^2 w^2
\end{matrix}\right)\;.
\end{eqnarray} 
Eq.~(\ref{eq:newG}) indicates that parameters $v,w,k$ can be estimated with the individual highest estimation precision (i.e. the Heisenberg scaling $1/T$) in the vicinity of TPT.
Since the condition of generating the TPT in this model refers to all the parameters, it enables us to extract the information about each parameter by virtue of the TPT.
Repeating the pertinent calculations with the Bloch vector $\tilde{r}_\text{in}$, according to Eq.~(\ref{eq:bc}) the  Berry curvature matrix is deduced as
\begin{eqnarray}\label{eq:SSHBC}
{\Omega}'=\bm{0}\;,
\end{eqnarray}
where any matrix element ${\Omega}'_{p q}=0$ since $(\vec{e}_p \times \vec{e}_{q}) \cdot \tilde{r}_\text{in}^\prime=0$ with $p,q=\{v,w,k\}$. 
The result of Eq.~(\ref{eq:SSHBC}) does not depend on the TPT condition, which differs from the circumstance indicated by  Eq.~(\ref{eq:newBC}) in the model of Sec.~\ref{Subsec:model1}.
The winding number indicates the number of times that the vector $\vec{\nu}$ of Eq.~(\ref{eq:vv}) encircles the origin in the momentum space, which is written as
\begin{eqnarray}\label{eq:WN}
W&=&\frac{1}{2\pi} \int_{-\pi}^\pi \partial_k \arctan \left(\frac{w \sin k}{v+w \cos k}\right) dk\nonumber\\
	&=&\frac{1-\text{sgn}(v-w)}{2}\;,
\end{eqnarray}
where  \text{sgn}($\bullet$) represents  a sign function.
Eq.~(\ref{eq:WN}) is plotted in Fig.~\ref{FigWN}.
Inserting Eq.~(\ref{eq:SSHBC}) into Eq.~(\ref{eq:MI}), we get the FOM matrix as
\begin{eqnarray}
{R}'=\bm{0}\;.
\end{eqnarray}
Thus the parameters $v,w,k$ can be estimated at the same time, but the individual highest estimation precision cannot be simultaneously reached without the help of TPT.
\begin{figure}[!h]
	\centering
	\includegraphics[width=0.35\textwidth]{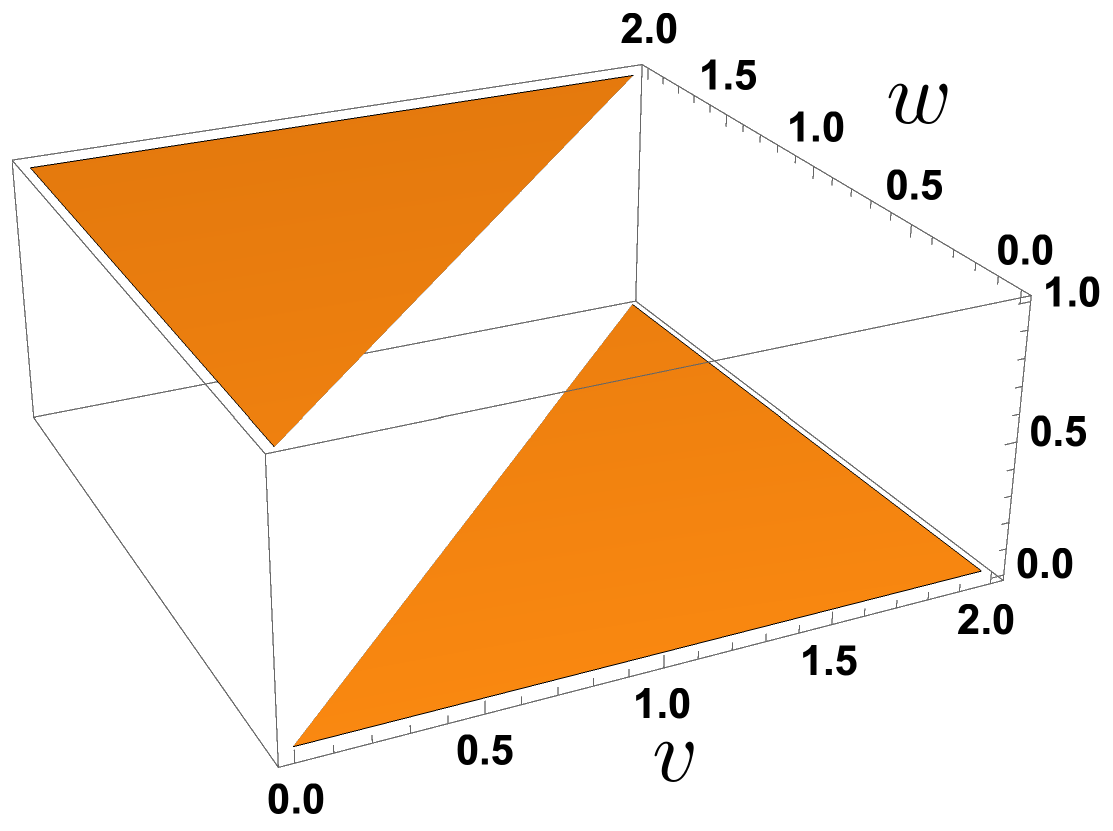}
	\caption
	{The relation among the winding number ${W}$ of  Eq.~(\ref{eq:WN}), the parameters $v$ and $w$ is plotted with the yellow surface.  The topological phase transition occurs when $v=w$, and $W=1$ ($W=0$) for  $v<w$ ($v>w$). }
	\label{FigWN}
\end{figure}

To avoid confusion, we remind  that Eqs.~(\ref{eq:thetanew})-(\ref{eq:gr}) and (\ref{eq:ncv})-(\ref{eq:nck}) correspond to an optimal input state maximizing the diagonal elements of the QFIM, whereas Eqs.~(\ref{eq:Gmatrix})-(\ref{eq:ave_BCmatrix}) and (\ref{eq:Gprime})-(\ref{eq:SSHBC}) correspond to an input state given by the ground state of the Hamiltonian.

\section{sensing approach based on TPT {VS} control-enhanced sensing approach}\label{Sec:VS}
To elucidate the benefits of utilizing the sensing approach based on the TPT, we still investigate the two SU(2) TPT models in Sec.~\ref{Sec:topo}. We show in this section that the TPT sensing protocol can reach the same highest estimation precision at the TPT point as the control-enhanced sensing protocol. 
Moreover, the experimental burden of the probe state preparation can be relaxed, since the probe state is only the ground state of the Hamiltonian rather than the entangled state that is necessary for the control-enhanced one.

Another approach for enhancing the estimation precision of an SU(2) parameterization process and eliminating the measurement incompatibility is to introduce the suitable quantum control optimizing dynamic systems~\cite{Yuan2016,Hou2021,Yang2022}. 
As shown in Fig.~\ref{Fig_scheme}(b), the probe state is now replaced with a two-qubit maximally entangled state $\hat{\rho}_\text{in}=|\psi_{SA}\rangle \langle \psi_{SA}|$, and an ancillary channel is introduced in the SU(2) dynamics system.
Quantum control reads
\begin{eqnarray}\label{eq:Hc}
	\hat{H}_c=\mathbf{X}_c \cdot \vec{J}\;,
\end{eqnarray}
where $\mathbf{X}_c=(X_1^{(c)}(\bm{ \tilde{\lambda}}),X_2^{(c)}(\bm{ \tilde{\lambda}}),X_3^{(c)}(\bm{ \tilde{\lambda}}))$ is a three-dimensional vector, $X_l^{(c)}(\bm{ \tilde{\lambda}})$ is a function of the estimated value $\bm{ \tilde{\lambda}}$ of  $\bm\lambda$,  $X_l^{(c)}(\bm \lambda)$ denotes a perfect quantum control.
In Fig.~\ref{Fig_scheme}(b) the $k$-th unitary cell ($k=\{1,2,\cdots,N\}$) can be expressed by $\hat{H}_k=\hat{H}(\bm\lambda)+\hat{H}_c=\mathbf{S} \cdot \vec{J}$ with a three-dimensional vector $\mathbf{S}=\mathbf{X}+\mathbf{X}_c$.
The optimal quantum control is proven to be~\cite{Yuan2016,Hou2021,Yang2022}
\begin{eqnarray}\label{eq:con}
	\hat{H}_c=-\hat{H}(\bm{\lambda})\;.
\end{eqnarray}
The gauge potential (\ref{eq:Hdef}) allows for a connection between the already obtained consequences in the control-enhanced quantum multiparameter estimation and the geometry of quantum state space.
Accordingly, the effectiveness of using quantum control to enhance parameter estimation precision and to get rid of the measurement incompatibility can also be proven by virtue of geometrical characteristics of the SU(2) model. The control-enhanced QMT associated with the parameters $\lambda_\mu$ and $\lambda_\nu$ is given by (see Appendix~\ref{Subsec:control})
\begin{eqnarray}
	{g}_{\mu \nu}^{(c)}=\frac{T^2}{4} (\partial_\mu \mathbf{X} \cdot \partial_\nu \mathbf{X})\;.
\end{eqnarray}
For $\lambda_\mu=\lambda_\nu$, it reduces to
\begin{eqnarray}\label{eq:gc}
	{g}_{\mu \mu}^{(c)}=\frac{T^2}{4}|\partial_\mu \mathbf{X}|^2\;.
\end{eqnarray} 
The  corresponding Berry curvature and the FOM are
\begin{eqnarray}
	{\Omega}_{\mu \nu}^{(c)}={r}_{\mu \nu}^{(c)}=0\;.
\end{eqnarray}

For the TPT model of Sec.~\ref{Subsec:model1}, we observed that the Hamiltonian (\ref{eq:TH}) under the TPT condition of $\theta=\pi, r=1$  reduces to a zero matrix.
This achieves the same consequence as employing quantum control of Eq.~(\ref{eq:con}) to optimize the original dynamics ($\hat{H}+\hat{H}_c=\mathbf{0}$).
Inserting Eq.~(\ref{eq:m}) into Eq.~(\ref{eq:gc}), one gets the QMTs as
	\begin{eqnarray}
		{g}_{\theta \theta}^{(c)}&=&\frac{T^2}{4}|\partial_\theta \vec{m}|^2
		= T^2\;,\label{eq:QMTtheta}\\
		{g}_{\phi \phi}^{(c)}&=&\frac{T^2}{4}|\partial_\phi \vec{m}|^2
		= T^2 \sin^2\theta\label{eq:QMTphi}\;,\\
		{g}_{r r}^{(c)}&=&\frac{T^2}{4}|\partial_r \vec{m}|^2=T^2\label{eq:QMTr}\;.
	\end{eqnarray}
Eqs.~(\ref{eq:QMTtheta})-(\ref{eq:QMTr}) are individually plotted in Figs.~\ref{Figc}(a)-(c) with the blue surfaces for the given value of $\theta=\pi$. 
Compared with the nonzero elements of Eqs.~(\ref{eq:LimitG}), (\ref{eq:QMTtheta})-(\ref{eq:QMTr}), we can see that the sensing approach based on the TPT can achieve the same highest estimation precision with the control-enhanced sensing approach, but the former has no requirement for the entangled probe state. This is exhibited in Fig.~\ref{Figc}(a) where the yellow and blue surfaces only intersect at the position of TPT.
Additionally, for the control-enhanced sensing strategy as shown in Eqs.~(\ref{eq:QMTtheta})-(\ref{eq:QMTr}), the parameters $\theta$, $\phi$, and $r$ can be simultaneously estimated for the case of $\theta \in (0,\pi)$. In contrast, for the TPT sensing strategy as shown in Eqs.~(\ref{eq:LimitG}), only the parameters $\theta$ and $r$ can be simultaneously estimated  with the individual highest estimation precision at the TPT point. The parameters associated with the TPT of the system are only a part of all the parameters ($\theta,r$) in this model.

Similar to the TPT model of Sec.~\ref{Subsec:model2}, the Hamiltonian (\ref{eq:SSH}) under the TPT condition of $v=w, k=\pi$ reduces to a zero matrix. This achieves the same consequence as employing the quantum control of Eq.~(\ref{eq:con}). 
By inserting Eq.~(\ref{eq:vv}) into Eq.~(\ref{eq:gc}), the QMTs are written as
	\begin{eqnarray}
		{g}_{v v}^{(c)}&=&\frac{T^2}{4}|\partial_v \vec{\nu}|^2=T^2\;,\label{eq:cv}\\
		{g}_{w w}^{(c)}&=&\frac{T^2}{4}|\partial_w \vec{\nu}|^2=T^2\;,\label{eq:cw}\\
		{g}_{k k}^{(c)}&=&\frac{T^2}{4}|\partial_k \vec{\nu}|^2=T^2 w^2\label{eq:ck}\;.
	\end{eqnarray}
Eqs.~(\ref{eq:cv})-(\ref{eq:ck}) are plotted in Figs.~\ref{FigSSH}(a)-(c) with the blue surfaces for the given value of $k=\pi$, respectively.
Compared with the nonzero elements of Eq.~(\ref{eq:newG}), (\ref{eq:cv})-(\ref{eq:ck}), we can see that the sensing approach based on the TPT can achieve the same highest estimation precision to the control-enhanced one, but the former has no requirement for the entangled probe state. 
However, the parameters $w$, $v$ and $k$ can be simultaneously estimated by two sensing approaches. This can be verified by comparing Eq.~(\ref{eq:newG}) to Eqs.~(\ref{eq:cv})-(\ref{eq:ck}). The parameters associated with the TPT of the system refer to all the to-be-estimated parameters ($w,v,k$) in this model.

\section{Adaptive multiparameter estimation strategy based on TPT}\label{Sec:scheme}
In this section, we use the fact that parameters have fixed values at the TPT point to estimate the initial values of  parameters associated with the TPT of the system.
In order to effectively estimate the initial values of  parameters associated with the TPT, here we propose an adaptive multiparameter estimation strategy.
The adaptive multiparameter estimation strategy is depicted in Fig.~1(c), and the TPT parameters are denoted by $\bm{\epsilon}=\{\epsilon_i\}$ ($i\in[1,3]$) for an SU(2) parameterization process. They  are usually  the subset of the parameters encoded in the Hamiltonian $\hat{H}(\bm{\lambda})$, i.e. $\bm{\epsilon} \in \bm{\lambda}$.
The initial values of $\bm{\epsilon}$ are unknown and to be estimated.  The TPT point  signals these parameters being the critical values, with which the quantum metric tensor (i.e. the QFI) presents a peak. 
In this way, we can employ the following adaptive TPT strategy to estimate the initial values of $\bm{\epsilon}$. 
We continuously adjust these parameters  from the initial points step by step until  the TPT point is approached.
According to the adjustment steps and the critical values, the initial values of $\bm{\epsilon}$ can be worked out. 
In the folloing we use two TPT models of Sec.~\ref{Sec:topo} as examples to show this adaptive estimation process.

Two dominant methods have been developed to experimentally measure the quantum metric tensor matrix (or the quantum Fisher information matrix): the periodic parameter modulation scheme~\cite{Ozawa2018}, and the suddenly quench scheme~\cite{Yu2019}. 
In the present investigation, we propose to check if the TPT point is reached by measuring the maximum of the quantum metric tensor.
Although our proposal works in principle, the recognition of the maximum of the QFIM may be a challenging work in the experiment because the change of the QFIM is small as the parameters approach the TPT point. Other methods should be considered. For example, one can identify the TPT point more clearly by measuring either Berry curvature~\cite{Cai2019} or the first Chern number~\cite{Cai2022}. Those quantities are also measurable in the experiment and have an abrupt change at the TPT point, as shown in Fig.~\ref{FigBC}(a) and Fig.~\ref{FigBC}(b).

\subsection{The canonical model}\label{subsec:model1}
For the model of Sec.~\ref{Subsec:model1}, we assume the initial values of unknown parameters to be $\{\theta_0,r_0\}$ for a given value of $\phi_0$.
The optimal probe state is the single-qubit pure state with the Bloch vector satisfying $\vec{e}_\theta \cdot \vec{r}_\text{in}=0$ ($||\vec{r}_\text{in}||=1$), which typically depends on the  values of  $\theta_0,\phi_0$ and $r_0$.
But we noticed that in the proximity of the TPT the unit vector $\vec{e}_\theta$ can be reduced to $\lim\limits_{\substack{\theta \to \pi \\	r \to 1}}\vec{e}_\theta=\{\cos \phi_0, \sin \phi_0,0\}$ (see Appendix~\ref{App:vector}). Thus the Bloch vector of the probe state can be simplified as $\vec{r}_\text{in}=\{c_x,-c_x/\tan\phi_0,c_z\}$ with the real numbers $c_x,c_z$.

One then alters the values of $\theta_0$ and $r_0$ step by step  until the obtainable estimation precision of $\theta_0$ reaches to the highest estimation precision (i.e. $g_{\theta \theta}^{(M)}$ of Eq.~(\ref{eq:thetanew}) approaches the ``peak" $T^2$),
in this moment  the TPT point is reached.
The number of times of altering $\theta_0$ ($r_0$) is recorded as $N_\theta$ ($N_r$), the controllable step is $\delta \theta_i$ ($\delta r_i$) for the $i$-th adjustment. Based on the feature in Fig.~\ref{Figc}(a), we have
	\begin{eqnarray}\label{eq:adaptive}
		\left\{\begin{matrix}
		\pi=\theta_0+\sum\limits_{i=1}^{N_\theta} \delta \theta_i\\\
		1=r_0+\sum\limits_{i=1}^{N_r}  \delta r_i
		\end{matrix}\right.\;.
	\end{eqnarray}
Thus as shown in  Eq.~(\ref{eq:adaptive}), the initial values of $\theta_0$ and $r_0$ can be derived from the gap between the TPT point and recorded total displacements.

As discussed in Sec.~\ref{Subsec:model1}, at the TPT point the QMT of $\theta_0$  is identical with the counterpart contributed by quantum control. Accordingly, the operation of driving the system to the TPT point can serve the same purpose  as employing quantum control $\hat{H}_c=-\hat{H}=-\vec{m} \cdot \vec{J}$, i.e. 
	\begin{eqnarray}\label{eq:HTPT}
		\hat{H}_\text{TPT}=\vec{n}\cdot \vec{J}=\hat{H}+\hat{H}_c\;,
	\end{eqnarray}
where 
	\begin{eqnarray}
		\vec{n}&=&2H_0\left(\sin \left(\theta_0+\sum\limits_{i=1}^{N_\theta} \delta \theta_i\right)  \cos\phi_0, \right.\nonumber\\
		&&\left.\sin \left(\theta_0+\sum\limits_{i=1}^{N_\theta} \delta \theta_i\right) \sin\phi_0,\right.\nonumber\\
		&&\left.\cos\left(\theta_0+\sum\limits_{i=1}^{N_\theta} \delta \theta_i\right)+\left(r_0+\sum\limits_{i=1}^{N_r}  \delta r_i\right)\right).
	\end{eqnarray}
Due to the experimental imperfections Eq.~(\ref{eq:HTPT}) is frequently not a zero matrix, yet it is still valid to estimate parameters effectively. In Appendix~\ref{App:sim}, the impact of these flaws on the estimation precision are numerically simulated.

\subsection{The SSH model}
For the SSH model of Sec.~\ref{Subsec:model2}, we assume the initial values of unknown parameters to be $\{k_0, v_0\}$ for a given value of $w_0$.
The optimal probe state is the single-qubit pure state with the Bloch vector satisfying $\vec{e}_k \cdot \tilde{r}_\text{in}=0$ ($||\tilde{r}_\text{in}||=1$),
which typically depends on the values of $k_0,w_0$ and $v_0$.
Since $\vec{e}_k$ can be reduced to  $\lim\limits_{\substack{k \to \pi \\ w \to v}}\vec{e}_k=\{0,1,0\}$ at the TPT point (see Appendix~\ref{App:vector}), 
the Bloch vector of the probe state can be simplified as $\tilde{r}_\text{in}=\{d_x,-1,d_z\}$ with the  real numbers $d_x,d_z$.

One then alters the values of $k_0$ and $v_0$ step by step until the obtainable estimation precision of $k_0$ reaches to the highest estimation precision (i.e. $g_{k k}^{(M)}$ of Eq.~(\ref{eq:nck}) approaches the ``peak" $T^2 w_0^2$),
in this moment the TPT point is reached.
The number of times of altering $k_0$ ($v_0$) is recorded as $N_k$ ($N_v$), the controllable step is $\delta k_i$ ($\delta v_i$) for $i$-th adjustment. 
Based on the feature in Fig.~\ref{FigSSH}(c), we have
\begin{eqnarray}\label{eq:Adap}
	\left\{\begin{matrix}
		\pi=k_0+\sum\limits_{i=1}^{N_k} \delta k_i\\\
		w_0=v_0+\sum\limits_{i=1}^{N_v}  \delta v_i
	\end{matrix}\right.\;.
\end{eqnarray}
Thus as shown in  Eq.~(\ref{eq:Adap}), the initial values of $k_0$ and $v_0$ can be derived from the gap between the TPT point and recorded total displacements.

At the TPT point the QMT of $k_0$ is equivalent to  the counterpart  contributed by quantum control $\hat{H}_c^\prime=-\hat{H}^\prime=-\vec{\nu} \cdot \vec{J}$, we therefore have
	\begin{eqnarray}\label{eq:Hwin}
		\hat{H}_\text{TPT}^\prime=\vec{u}\cdot \vec{J}=\hat{H}^\prime+\hat{H}_c^\prime\;,
	\end{eqnarray}
	where 
	\begin{eqnarray}
		\vec{u}&=&2\left(\left(v_0+\sum\limits_{i=1}^{N_v}  \delta v_i\right)+w_0\cos \left(k_0+\sum\limits_{i=1}^{N_k} \delta k_i\right), \right.\nonumber\\
		&&\left.w_0 \sin \left(k_0+\sum\limits_{i=1}^{N_k} \delta k_i\right),0\right)\;.
	\end{eqnarray}
Theoretically Eq.~(\ref{eq:Hwin}) is a zero matrix, but it will be affected by the experimental imperfections, one can do simulations as those in Appendix~\ref{App:sim} to analyze the impacts of these imperfections on the estimation precision.

In summary, the TPT estimation strategy can reach the Heisenberg scaling by employing the ground state of the Hamiltonian as the probe state rather than the entangled state which required in the control-enhanced one. Besides, in the TPT strategy we use the peculiarity that parameters have fixed values at the TPT point to estimate the unknown initial values of  parameters associated with the TPT of the system.
These are the advantages of the TPT strategy.
The disadvantage of this TPT strategy is that only the parameters associated with the TPT of the system can be simultaneously estimated with the individual highest estimation precision. In contrast, the control-enhanced one gives the possibility of simultaneous optimal estimation for all the parameters.

\section{Summary}\label{Sec:dissum}
We have deduced the geometrical properties of quantum states that are encoded in an SU(2) dynamics system, including the QGT, QMT, Berry curvature, and the first Chern number.
These geometrical quantities can be experimentally measured by the periodic parameter modulation scheme~\cite{Klees2020,Ozawa2018} or the suddenly quench scheme~\cite{Yu2019}.
By examining the two SU(2) TPT models, we have revealed that multiple parameters that drive the system to the TPT can be simultaneously estimated with the individual highest precision around the TPT point.
We have also discovered that the proposed TPT sensing protocol can achieve the same metrology performance as the control-enhanced sensing protocol. 
Moreover, the probe state of the present scheme is the ground state of the Hamiltonian rather than the entangled state that is necessary for the control-enhanced sensing protocol.
This may effectively relax the experimental burden of the probe state preparation.
In addition, an adaptive multiparameter estimation strategy has been suggested and applied to the two SU(2) TPT models.
The gradient ascent pulse engineering (GRAPE) and the machine learning (ML) algorithm~\cite{Adaptive2021} can be utilized to further speed up this estimation process.
The connection between quantum phase transition (QPT) in quantum many-body systems and quantum metrology has been studied in recent years~\cite{Chu2023,Ying2022,critical,Garbe2020,WeipingZhang2023,Yin2019}.
However,  our work might be categorized as quantum critical metrology with the employment of TPT resources.

\begin{acknowledgments}
We thank Zhibo Hou from USTC for helpful comments. This research is supported by the National Natural Science Foundation of China (NSFC) (Grants No. 12204371, 12074307), Shaanxi Fundamental Science Research Project for Mathematics and Physics (Grant No. 22JSZ004), and Shaanxi Natural Science Basic Research Program (Grant No. 2021JQ-008).
\end{acknowledgments}

\appendix
\begin{widetext}
\section{Figure of merit of measurement incompatibility}\label{App:incertainty_relation}
The Robertson-Schr$\ddot{o}$dinger uncertainty relation with respect to two Hermitian operators $\hat{A}$, $\hat{B}$ is
\begin{eqnarray}\label{eq:relation}
\langle (\Delta \hat{A})^2\rangle \langle (\Delta \hat{B})^2 \rangle \ge \frac{1}{4}|\langle [\hat{A},\hat{B}] \rangle|^2+\frac{1}{4} \langle\{ \Delta \hat{A},\Delta \hat{B} \}\rangle^2, 
\end{eqnarray}
where $\Delta \hat{A}(\hat{B})=\hat{A}(\hat{B})-\langle \hat{A}(\hat{B})\rangle \hat{I}$ with an identity operator $\hat{I}$. 
Substituting the gauge potentials $\tilde{\mathcal{A}}_{\mu}$, $\tilde{\mathcal{A}}_{\nu}$ into Eq.~(\ref{eq:relation}) with the initial probe state $|\psi\rangle$, we get
\begin{eqnarray}\label{eq:A2}
\langle  \psi| (\Delta \tilde{\mathcal{A}}_{\mu})^2 |\psi \rangle  \langle \psi| (\Delta \tilde{\mathcal{A}}_{\nu})^2 |\psi \rangle
- \frac{1}{4}\langle\psi| \{ \Delta \tilde{\mathcal{A}}_{\mu}, \Delta \tilde{\mathcal{A}}_{\nu} \} |\psi\rangle^2
\ge \frac{1}{4} |\langle \psi| [\tilde{\mathcal{A}}_{\mu},\tilde{\mathcal{A}}_{\nu}] |\psi\rangle |^2\;.
\end{eqnarray}
Using Eqs.~(\ref{eq:QM}) and (\ref{eq:BCC}) to rewrite Eq.~(\ref{eq:A2}), one has
\begin{eqnarray}\label{eq:inequality}
g_{\mu \mu} g_{\nu \nu} -g_{\mu \nu}^2 \ge \frac{1}{4} \Omega_{\mu \nu}^2 \ge 0\;,
\end{eqnarray}
where the left-side is exactly the determinant of $\mathcal{G}_{\mu \nu}$ of Eq.~(\ref{eq:g2}). 
The FOM  based on Eq.~(\ref{eq:inequality})  can be defined as
\begin{eqnarray}\label{eq:FOM}
r_{\mu \nu}=\frac{\Omega_{\mu \nu}}{2\sqrt{\text{Det}\left[\mathcal{G}_{\mu \nu} \right]}} \in [0,1]\;.
\end{eqnarray}
Since the Berry curvature $\Omega_{\mu \nu}=0$ is equivalent to the weak commutation condition~\cite{Carollo2019,Yang2022}, $r_{\mu \nu}=0$ means that two parameters $\lambda_\mu$, $\lambda_\nu$ can be simultaneously estimated.
However, $r_{\mu \nu}=1$ indicates that the estimation precision of two parameters are maximally exclusive.
The upper bound of the difference between the HCRB and QCRB in Ref.~\cite{Carollo2019} and the Branciard uncertainty relation in Ref.~\cite{Lu2021}  both make reference to this FOM definition.

\section{Effectiveness of quantum control in multiparameter estimation}\label{App:control}
\subsection{Without quantum control} \label{Subsec:without}
In the multiparameter estimation scheme of Fig.~\ref{Fig_scheme}(a), 
the initial probe state is assumed to be a single-qubit pure state $\hat{\rho}_\text{in}=\hat{I}/2 +\vec{r}_\text{in} \cdot \vec{J}$ with the Bloch vector $\vec{r}_\text{in}$.
The pivotal Hermitian operator studied in Ref.~\cite{Yang2022} (see Eq.~(3) of Ref.~\cite{Yang2022}) is exactly the negative gauge potential, 
therefore the gauge potential  can be written as
\begin{eqnarray}\label{eq:GPwo}
\tilde{\mathcal{A}}_{\ell}=-|\mathbf{Y}_\ell|\vec{e}_\ell \cdot \vec{J}\;,
\end{eqnarray}
with an unit vector
	\begin{eqnarray}\label{eq:unitvector}
		\vec{e}_\ell=\frac{1}{|\mathbf{Y}_\ell|} \left\lbrace 
		-T ( \partial_\ell \mathbf{X})+\frac{|\partial_\ell \mathbf{X}||\sin \alpha_\ell|}{|\mathbf{X}|}
		\Big\{[ \sin(T|\mathbf{X}|)-T|\mathbf{X}|] \vec{v}_{\ell,2} +\left[1-\cos(T|\mathbf{X}|) \right]  \vec{v}_{\ell,1}
		\Big\} \right\rbrace  \;,
	\end{eqnarray}
	where $T=t N$ is the total evolution time,  
	$\alpha_\ell:=\left<\mathbf{X},\partial_\ell \mathbf{X}\right>$ represents the angle between vectors $\mathbf{X}$ and $\partial_\ell \mathbf{X}$,  $\vec{v}_{\ell,1}=\frac{\mathbf{X} \times \partial_\ell \mathbf{X}}{\left| \mathbf{X}\right| \left| \partial_\ell \mathbf{X}\right| \sin \alpha_\ell }$, $  \vec{v}_{\ell,2}=\frac{\mathbf{X} \times
		\left(\mathbf{X} \times \partial_\ell \mathbf{X}\right)}{\left| \mathbf{X}\right|^2 \left| \partial_\ell \mathbf{X}\right| | \sin \alpha_\ell|}$ and
	\begin{eqnarray}\label{eq:Ymodul}
		|\mathbf{Y}_\ell|=\sqrt{T^2 |\partial_\ell \mathbf{X} |^2 \cos^2\alpha_\ell+\frac{4|\partial_\ell \mathbf{X}|^2 \sin^2 \alpha_\ell}{|\mathbf{X}|^2} \sin^2 \left(\frac{T|\mathbf{X}|}{2} \right)}\;.
	\end{eqnarray}
We especially noticed that
	\begin{eqnarray}\label{eq:upper}
		0 \le  |\mathbf{Y}_\ell|^2=T^2 |\partial_\ell \mathbf{X}|^2 \left[\cos^2 \alpha_\ell+\sin^2 \alpha_\ell \left(\frac{\sin\left(\frac{T|\mathbf{X}|}{2}\right)}{\frac{T|\mathbf{X}|}{2}}\right)^2\right]\le T^2 |\partial_\ell \mathbf{X}|^2\;.
\end{eqnarray}

Inserting Eq.~(\ref{eq:GPwo}) into Eq.~(\ref{eq:AAnew}), we have the quantum geometric tensor 
\begin{eqnarray}\label{eq:QGT}
\chi_{\mu \nu}=\frac{|\mathbf{Y}_\mu| |\mathbf{Y}_\nu|}{4} \left[(\vec{e}_\mu \cdot \vec{e}_\nu)-(\vec{e}_\mu \cdot \vec{r}_\text{in}) (\vec{e}_\nu \cdot \vec{r}_\text{in})+i (\vec{e}_\mu \times \vec{e}_\nu) \cdot \vec{r}_\text{in}
\right],
\end{eqnarray}
where
$\text{Tr}\left[\tilde{\mathcal{A}}_\mu \tilde{\mathcal{A}}_{\nu} \hat{\rho}_\text{in}\right]=\frac{|\mathbf{Y}_\mu| |\mathbf{Y}_\nu|}{4} \left[ (\vec{e}_\mu \cdot \vec{e}_\nu)+i (\vec{e}_\mu \times \vec{e}_\nu) \cdot \vec{r}_\text{in}\right], 
\text{Tr}\left[ \tilde{\mathcal{A}}_\mu \hat{\rho}_\text{in} \right]=\frac{-|\mathbf{Y}_\mu| }{2} (\vec{e}_\mu \cdot \vec{r}_\text{in})$ are figured out.
Thus the quantum metric tensor can be expressed by
\begin{eqnarray}\label{eq:g}
	g_{\mu \nu}=\frac{|\mathbf{Y}_\mu||\mathbf{Y}_\nu|}{4} [(\vec{e}_\mu \cdot \vec{e}_{\nu})-(\vec{e}_\mu \cdot \vec{r}_\text{in})(\vec{e}_{\nu} \cdot \vec{r}_\text{in})]\;.
\end{eqnarray}
For $\lambda_\mu=\lambda_\nu$, it reduces to
\begin{eqnarray}\label{eq:gdia}
	g_{\mu \mu}=\frac{|\mathbf{Y}_\mu|^2}{4} [1-(\vec{e}_\mu \cdot \vec{r}_\text{in})^2]\;.
\end{eqnarray}
The corresponding Berry curvature and the first Chern number are 
\begin{eqnarray}
	\Omega_{\mu \nu}&=&-\frac{|\mathbf{Y}_\mu| |\mathbf{Y}_\nu|}{2} (\vec{e}_\mu \times \vec{e}_\nu) \cdot \vec{r}_\text{in}\;,\label{eq:Berry_curvature}\\
	C_{\mu \nu}&=&\frac{1}{2\pi}\int_{S^2}\Omega_{\mu \nu} d\lambda_\mu \wedge d\lambda_\nu
	=\frac{-1}{4\pi} \int_{S^2} |\mathbf{Y}_\mu| |\mathbf{Y}_{\nu}| (\vec{e}_\mu \times \vec{e}_{\nu}) \cdot \vec{r}_\text{in} d\lambda_\mu \wedge d\lambda_{\nu},\label{eq:Chern}
\end{eqnarray}
where $\lambda_\mu, \lambda_\nu \in S^2$ (Bloch sphere). 
By substituting Eqs.~(\ref{eq:g})-(\ref{eq:Berry_curvature}) into Eq.~(\ref{eq:MI}), under the condition of $\vec{e}_\mu \cdot \vec{r}_\text{in}=\vec{e}_\nu \cdot \vec{r}_\text{in}=0$ the FOM is renewed as
\begin{eqnarray}\label{eq:charac}
	r_{\mu \nu}= \frac{-(\vec{e}_\mu \times \vec{e}_\nu)\cdot \vec{r}_\text{in}}{\sqrt{1-(\vec{e}_\mu \cdot \vec{e}_\nu)^2}}\;,
\end{eqnarray}
where
$\text{Det}[\mathcal{G}_{\mu\nu}]=|\mathbf{Y}_\mu|^2 |\mathbf{Y}_\nu|^2 [1-(\vec{e}_\mu \cdot \vec{e}_\nu)^2]/16$ is employed.
To achieve the simultaneous estimation for $\lambda_\mu$ and $\lambda_\nu$ with the individual highest precision,  Eqs.~(\ref{eq:gdia}) and (\ref{eq:charac}) require that
\begin{eqnarray}\label{eq:condition}
	\left\{\begin{matrix}
		\vec{e}_\mu \cdot \vec{r}_\text{in}&=&0\\
		\vec{e}_\nu \cdot \vec{r}_\text{in}&=&0\\
		(\vec{e}_\mu \times \vec{e}_\nu) \cdot \vec{r}_\text{in}&=&0
	\end{matrix}\right.\;,
\end{eqnarray}
should be satisfied at the same time, but obviously it is unable to find such Bloch vector $\vec{r}_\text{in}$.
Accordingly, the simultaneous optimal estimation of multiple parameter cannot be fulfilled in this case.

\subsection{With quantum control}\label{Subsec:control}
In the case of quantum control, $\hat{H}_c=-\hat{H}(\bm \lambda)$ (or $\mathbf{X}_c=-\mathbf{X}$) has already been proven to be an effective control form in Refs.~\cite{Hou2021,Yang2022}.
As shown in Fig.~\ref{Fig_scheme}(b), the probe state is now replaced with a two-qubit maximally entangled state $\hat{\rho}_\text{in}=|\psi_{SA}\rangle \langle \psi_{SA}|$, and an ancillary channel is introduced in the SU(2) dynamics system.
According to Ref.~\cite{Yang2022}, we get the gauge potential 
\begin{eqnarray}\label{eq:GPwoNew}
\tilde{\mathcal{A}}_{\ell}^{(c)}=T (\partial_{\ell} \mathbf{X} \cdot \vec{J}) \otimes \hat{I}\;,
\end{eqnarray} 
where $\hat{I}$ is an identity operator. 
Then the following hints are employed
\begin{eqnarray}\label{eq:part1}
\text{Tr}\left[\tilde{\mathcal{A}}_\mu \tilde{\mathcal{A}}_\nu \hat{\rho}_\text{in}  \right]&=&\text{Tr}\left[
\left(T(\partial_\mu \mathbf{X} \cdot \vec{J}) \otimes \hat{I}\right) \left(T (\partial_\nu \mathbf{X} \cdot \vec{J})  \otimes \hat{I}  \right) \hat{\rho}_\text{in}\right]\nonumber\\
&=&T^2 \text{Tr} \left[\left((\partial_\mu \mathbf{X} \cdot \vec{J}) (\partial_\nu \mathbf{X} \cdot \vec{J}) \otimes \hat{I}\right) \hat{\rho}_\text{in}\right]\nonumber\\
&=& T^2 \text{Tr}\left[\hat{\rho}_s \left(\frac{1}{4} (\partial_\mu \mathbf{X} \cdot \partial_\nu \mathbf{X})\hat{I}+\frac{i}{2} (\partial_\mu \mathbf{X} \times \partial_\nu \mathbf{X}) \cdot \vec{J}\right)\right] \nonumber\\
&=& T^2 \left\{ \frac{\partial_\mu \mathbf{X} \cdot \partial_\nu \mathbf{X}}{4} \text{Tr}[\hat{\rho}_s]
+\frac{i}{2} \text{Tr}\left[\hat{\rho}_s \left((\partial_\mu \mathbf{X} \times \partial_\nu \mathbf{X})\cdot \vec{J}\right)\right]\right\}\;,
\end{eqnarray}
where $\hat{\rho}_s$ is the reduced density operator of $\hat{\rho}_\text{in}=|\psi_{SA}\rangle \langle \psi_{SA}|$ after tracing out the ancillary part, i.e. $\hat{\rho}_s=\text{Tr}_A\left[\hat{\rho}_\text{in} \right]=\hat{I}/2$. We diagonize the matrix $(\partial_\mu \mathbf{X} \times \partial_\nu \mathbf{X}) \cdot \vec{J}$ as
\begin{eqnarray}\label{eq:d}
(\partial_\mu \mathbf{X} \times \partial_\nu \mathbf{X}) \cdot \vec{J}=\hat{Y} \left( \begin{matrix}
+a|\partial_\mu \mathbf{X} \times \partial_\nu \mathbf{X}|&0\\
0&-a|\partial_\mu \mathbf{X} \times \partial_\nu \mathbf{X}|
\end{matrix}\right) \hat{Y}^\dagger\;,
\end{eqnarray}
where $\hat{Y}$ is an unitary matrix and $\pm a$ is the maximal (minimal) eigenvalue of $\hat{j}_m$ for $m=\{1,2,3\}$.
Plugging Eq.~(\ref{eq:d}) into Eq.~(\ref{eq:part1}) we have
\begin{eqnarray}\label{eq:t1}
\text{Tr}\left[\tilde{\mathcal{A}}_\mu \tilde{\mathcal{A}}_\nu \hat{\rho}_\text{in}  \right]&=&T^2 \left\{ \frac{\partial_\mu \mathbf{X} \cdot \partial_\nu \mathbf{X}}{4} \text{Tr}[\hat{\rho}_s]+
\frac{i}{2} \text{Tr}\left[\hat{Y}^\dagger \hat{\rho}_s \hat{Y} \left( \begin{matrix}
+a|\partial_\mu \mathbf{X}\times \partial_\nu \mathbf{X}|&0\\
0&-a|\partial_\mu \mathbf{X}\times \partial_\nu \mathbf{X}|
\end{matrix}\right)
\right]
\right\}=\frac{T^2}{4}(\partial_\mu \mathbf{X} \cdot \partial_\nu \mathbf{X}), \nonumber\\
\end{eqnarray}
where $\text{Tr}\left[\hat{\rho}_s\right]=1$ and $\hat{Y}^\dagger \hat{\rho}_s \hat{Y}=\hat{I}/2$.
Similarly, we have
\begin{eqnarray} \label{eq:t2}
\text{Tr}[\tilde{\mathcal{A}}_\ell \hat{\rho}_\text{in}]=\text{Tr}\left[(T(\partial_\ell \mathbf{X}\cdot \vec{J}) \otimes \hat{I}) \hat{\rho}_\text{in} 
\right]=T\; \text{Tr}\left[ \hat{\rho}_s (\partial_\ell \mathbf{X} \cdot \vec{J})\right]=T\; \text{Tr}\left[\hat{Z}^\dagger \hat{\rho}_s \hat{Z}\left( \begin{matrix}
+a |\partial_\ell \mathbf{X}| & 0\\
0& -a |\partial_\ell \mathbf{X}|
\end{matrix}\right)
\right]=0\;.
\end{eqnarray}
The diagonalization 
\begin{eqnarray}
 \partial_\ell \mathbf{X} \cdot \vec{J}=\hat{Z} \left(\begin{matrix}
+a |\partial_\ell \mathbf{X}| & 0\\
0& -a |\partial_\ell \mathbf{X}|
\end{matrix}\right)\hat{Z}^\dagger\;,
\end{eqnarray}
is operated with a unitary matrix $\hat{Z}$ and  $\hat{Z}^\dagger \hat{\rho}_s \hat{Z}=\hat{I}/2$.

Inserting Eqs.~(\ref{eq:t1}) and (\ref{eq:t2}) into (\ref{eq:AAnew}), we get the quantum geometric tensor  
\begin{eqnarray}
\chi_{\mu \nu}^{(c)}=\frac{T^2}{4}\left( \partial_{\mu} \mathbf{X} \cdot \partial_{\nu} \mathbf{X}  \right)\;,
\end{eqnarray}
where $T=tN$ is the total evolution time.
The quantum metric tensor therefore is
\begin{eqnarray}
	{g}_{\mu \nu}^{(c)}=\frac{T^2}{4} (\partial_\mu \mathbf{X} \cdot \partial_\nu \mathbf{X})\;.
\end{eqnarray}
For $\lambda_\mu=\lambda_\nu$, it reduces to
\begin{eqnarray}\label{eq:QMTnew}
	{g}_{\mu \mu}^{(c)}=\frac{T^2}{4}|\partial_\mu \mathbf{X}|^2\;.
\end{eqnarray} 
The  corresponding Berry curvature and the  first Chern number  are
\begin{eqnarray}
	{\Omega}_{\mu \nu}^{(c)}={C}^{(c)}_{\mu \nu}=0\;.\label{eq:BCnew}
\end{eqnarray}
The FOM of Eq.~(\ref{eq:MI}) yields
\begin{eqnarray}\label{eq:rnew}
	{r}_{\mu \nu}^{(c)}=0\;.
\end{eqnarray}
We demonstrated that, by virtue of quantum control, it is possible to achieve the individual highest estimation precision for multiple parameters by comparing to Eqs.~(\ref{eq:upper}), (\ref{eq:gdia}), and (\ref{eq:QMTnew}). 
Eq.~(\ref{eq:rnew}) indicates that the measurement incompatibility existed in the case of no control (see Eq.~(\ref{eq:condition}) can be avoided.

\section{Clarifications for simulation results}\label{App:comparison}
For a generic SU(2)  dynamics system with the Hamiltonian $H=\vec{d}(\bm \lambda)\cdot \hat{\sigma}$,
where $\vec{d}(\bm \lambda)=(d_1(\bm \lambda),d_2(\bm \lambda),d_3(\bm \lambda))$ is a three-dimensional vector and $\hat{\sigma}=\{\hat{\sigma}_x,\hat{\sigma}_y,\hat{\sigma}_z\}$ denotes the Pauli matrices, 
the Schmidt decomposition is written as 
\begin{eqnarray}\label{eq:g1}
	H=d(\bm \lambda)|g(\bm \lambda)\rangle \langle g(\bm \lambda)|-d(\bm \lambda)|e(\bm \lambda)\rangle \langle e(\bm \lambda)|\;,
\end{eqnarray}
with
\begin{eqnarray}
	d(\bm \lambda)&=&|\vec{d}(\bm \lambda)|, 
	|g(\bm \lambda)\rangle=\left(\begin{matrix}
		\cos(\theta/2)\\ e^{i\phi} \sin(\theta/2)
	\end{matrix}\right),
	 | e(\bm \lambda)\rangle=\left(\begin{matrix}
		\sin(\theta/2)\\ -e^{i\phi} \cos(\theta/2)
	\end{matrix}\right),
	\cos(\theta)=\frac{d_3(\bm \lambda)}{d(\bm \lambda)}\;,  e^{i\phi}=\frac{d_1(\bm \lambda)+i d_2(\bm \lambda)}{\sqrt{d_1^2(\bm \lambda)+d_2^2(\bm \lambda)}},
\end{eqnarray} 
for $\theta \in [0,\pi]$, $\phi \in [0,2\pi]$.
To investigate the relation between our results (see Eqs.~(\ref{eq:BCmatrix})-(\ref{eq:aveChern}) in the maintext) and the previous results~\cite{Cai2022,Cai2019,Goldman2018}, we further give the following discussions for the canonical TPT model of Sec.~\ref{Subsec:model1}.
In Refs.~{\cite{Cai2022,Cai2019,Goldman2018}} the encoded state $\hat{\rho}_\text{out}$ used to calculate the QGT of Eq.~(\ref{QGT}) is the ground state $|{g}(\theta,\phi,r)\rangle$ of the Hamiltonian (\ref{eq:TH}), i.e.
\begin{eqnarray}
	H=H_0 \left( \begin{matrix}\label{eq:e1}
		\cos\theta+r & \sin\theta e^{-i\phi}\\
		\sin\theta e^{i\phi} & -\cos\theta-r
	\end{matrix}\right) =H_0 |g(\theta,\phi,r)\rangle \langle g(\theta,\phi,r)|-H_0 |e(\theta,\phi,r)\rangle \langle e(\theta,\phi,r)|\;,
\end{eqnarray}
with
\begin{eqnarray}
	d(\theta,\phi,r)=H_0\sqrt{1+r^2+2r \cos\theta}\;, \quad |g(\theta,\phi,r)\rangle=\left(\begin{matrix}
		\cos(\theta'/2)\\ e^{i\phi} \sin(\theta'/2)
	\end{matrix}\right)\;,
	\quad | e( \theta,\phi,r)\rangle=\left(\begin{matrix}
		\sin(\theta'/2)\\ -e^{i\phi} \cos(\theta'/2)
	\end{matrix}\right)\;,
\end{eqnarray}
where $\theta'=\arccos((\cos\theta+r)/\sqrt{1+r^2+2r \cos\theta})$.
By reversing the unitary evolution, we get the initial probe state 
\begin{eqnarray}
	\hat{\rho}_\text{in}&=&\hat{U}(\theta,\phi,r)\hat{\rho}_\text{out}\hat{U}^\dagger(\theta,\phi,r)
	=e^{-iT H}|g(\theta,\phi,r)\rangle \langle g(\theta,\phi,r)|e^{iTH}\nonumber\\
	&=&\left(e^{-iTH_0}|g(\theta,\phi,r)\rangle \langle g(\theta,\phi,r)|+e^{iTH_0}|e(\theta,\phi,r)\rangle \langle e(\theta,\phi,r)|\right) |g(\theta,\phi,r)\rangle \langle g(\theta,\phi,r)|\nonumber\\
	&\times&\left(e^{iTH_0}|g(\theta,\phi,r)\rangle \langle g(\theta,\phi,r)|+e^{-iTH_0} |e(\theta,\phi,r)\rangle \langle e(\theta,\phi,r)| \right)\nonumber\\
	&=&|g(\theta,\phi,r)\rangle \langle g(\theta,\phi,r)|\;.
\end{eqnarray}
The aforementioned  probe state has the same expression as the encoded state, but $\theta,\phi,r$ involved in $\hat{\rho}_\text{in}$ are not variables but quantities that are the same as true values of the to-be-estimated parameters ${\theta,\phi,r}$.
Thus this probe state should be adaptively updated with the estimated values of $\theta,\phi,r$.
Differently, in our work the encoded state is 
\begin{eqnarray}
	|\tilde{\psi}(\theta,\phi,r)\rangle&=&\hat{U}(\theta,\phi,r)|g(\theta,\phi,r)\rangle=e^{-iTH}|g(\theta,\phi,r)\rangle\nonumber\\
	&=&\left(e^{-iTH_0}|g(\theta,\phi,r)\rangle \langle g(\theta,\phi,r)|+e^{iTH_0}|e(\theta,\phi,r)\rangle \langle e(\theta,\phi,r)|\right) |g(\theta,\phi,r)\rangle\nonumber\\
	&=&e^{-iTH_0}|g(\theta,\phi,r)\rangle\;.
\end{eqnarray}
Our encoded state has a difference $e^{-iTH_0}$ from the previous encoded state.
It follows that the coarse-grained Berry curvature (see the yellow curve of Fig.~\ref{FigBC}(a)) and the coarse-grained first Chern number (see the yellow curve of Fig.~\ref{FigBC}(b)) are twifold of the results in Refs.~\cite{Cai2022,Cai2019,Goldman2018}.

\section{Optimal measurement scheme for parameters $\{\theta, r\}$}\label{App:meas}
We need to determine if the weak commutation condition with regard to parameters $\theta$, $r$ can be satisfied before studying the optimal measurement scheme.  
The weak commutation condition~\cite{Yang2022} associated with the gauge potential writes
\begin{eqnarray}\label{eq:weakcom}
	\text{Tr}\left[[\tilde{\mathcal{A}}_{\theta},\tilde{\mathcal{A}}_{r}]\hat{\rho}_\text{in}\right]=
	\frac{i |\mathbf{Y}_{\theta}| |\mathbf{Y}_{r}|}{2} (\vec{e}_\theta \times \vec{e}_r) \cdot \vec{r}_\text{in}\;,
\end{eqnarray}
where $\vec{e}_\theta$, $\vec{e}_r$ are given by Eq.~(\ref{eq:unitvector}).
Given the Bloch vector  $\vec{r}_\text{in}=(\sin\theta' \cos\phi,\sin\theta' \sin\phi,\cos\theta')$ with $\theta'=\arccos\left[{(\cos\theta +r)/}{\sqrt{1+r^2+2r \cos\theta}}\right]$ for the initial probe state $\hat{\rho}_\text{in}=|\psi(\theta,\phi,r)\rangle \langle \psi(\theta,\phi,r)|$,
Eq.~(\ref{eq:weakcom}) can be zero since $(\vec{e}_\theta \times \vec{e}_r) \cdot \vec{r}_\text{in}=0$.
Accordingly, it is always possible to reach the QCRB with respect to the parameters $\theta$ and $r$ if employing an optimal measurement scheme.
We emphasize that the parameters $\theta,\phi,r$ are just quantities in the initial probe state $|\psi(\theta,\phi,r)\rangle$, while they are variables in the encoded state $|\tilde{\psi}(\theta,\phi,r)\rangle$.

We now apply the optimal measurement scheme proposed in [Phys. Rev. Lett.  119, 130504 (2017)]  to our scenario.
The measurement scheme is constructed by a set of projectors $\{|\Upsilon_k \rangle \langle \Upsilon_k|\}$ for $k=\{1,2,3\}$ and
\begin{eqnarray}
	|{\Upsilon}_1\rangle&=&|\tilde{\psi}\rangle\;,\\
	|{\Upsilon}_2\rangle&=&|\omega_\theta \rangle=|\partial_\theta \tilde{\psi}\rangle\;,\label{eq:ot2}\\
	|{\Upsilon}_3\rangle&=&|\omega_r\rangle-\frac{\langle \omega_r|\omega_\theta \rangle}{\langle \omega_\theta|\omega_\theta \rangle}|\omega_\theta\rangle=|\partial_r \tilde{\psi}\rangle -\frac{\sin\theta}{1+r \cos\theta} |\partial_\theta \tilde{\psi}\rangle\;,\label{eq:ot3}
\end{eqnarray}
where Eqs.~(\ref{eq:ot2})-(\ref{eq:ot3}) are deduced via the Gram-Schmidt process, $|\omega_\ell\rangle=|\partial_\ell \tilde{\psi}\rangle+|\tilde{\psi}\rangle \langle \partial_\ell \tilde{\psi}|\tilde{\psi}\rangle$ for $\ell=\{\theta,r\}$ and $|\tilde{\psi}\rangle:=|\tilde{\psi}(\theta,\phi,r)\rangle=\cos\left(\frac{\theta'}{2}\right)|0\rangle+\sin\left(\frac{\theta'}{2}\right) e^{i\phi}|1\rangle$.
The  classical Fisher information matrix (CFIM) reads
\begin{eqnarray}\label{eq:CFIM}
	[\mathbf{J}]_{\ell,m}&=&\sum_{k=1}^3 \frac{\partial_\ell P(k|\lambda) \partial_{m}P(k|\lambda)}{P(k|\lambda)}\;,
\end{eqnarray}
where $P(k|\lambda)=\langle \tilde{\psi}|{\Upsilon}_k\rangle \langle {\Upsilon}_k |\tilde{\psi}\rangle$ and 
$\partial_\ell P(k|\lambda)=2\text{Re}[\langle \partial_\ell \tilde{\psi}|{\Upsilon}_k\rangle \langle {\Upsilon}_k|\tilde{\psi}\rangle ]$.
The QFIM writes
\begin{eqnarray}\label{eq:QFIM}
	[\mathbf{F}]_{\ell,\ell'}&=&4\text{Re}[\langle \partial_\ell \tilde{\psi}|\partial_{\ell'} \tilde{\psi}\rangle]+4\langle \partial_\ell \tilde{\psi}|\tilde{\psi}\rangle \langle \partial_{\ell'} \tilde{\psi}|\tilde{\psi}\rangle\;,
\end{eqnarray}
for $\ell,\ell'=\{\theta,r\}$.
One can prove that Eq.~(\ref{eq:CFIM}) equals to  Eq.~(\ref{eq:QFIM})  if and only if 
\begin{eqnarray}
	\text{lim}_{\varphi \to \lambda} \frac{\text{Im}[\langle \partial_\ell \tilde{\psi}| {\Upsilon}_k\rangle \langle \Upsilon_k| \tilde{\psi}\rangle ]}{|\langle {\Upsilon}_k|\tilde{\psi}\rangle|}=0\;,
\end{eqnarray}
for all $k \neq 1$.

\section{Limitation expressions of vectors}\label{App:vector}
For the canonical model of Eq.~(\ref{eq:TH}), we noticed that in the proximity of the TPT the unit vector $\vec{e}_\theta$ associated with the initial values $\{\theta_0,\phi_0,r_0\}$ can be reduced to 
		\begin{eqnarray}
			\lim\limits_{\substack{\theta \to \pi \\	r \to 1}}\vec{e}_\theta&=&\frac{1}{\mathcal{E}}\Big\{-2T\cos\theta_0\cos\phi_0 \xi^2+(1+r_0 \cos\theta_0)[-2\xi \sin\phi_0 \sin^2(T\sqrt{\xi})+(r_0+\cos\theta_0)\cos\phi_0(2T\xi-\sqrt{\xi}\sin(2T\sqrt{\xi}))],\nonumber\\
			&&-2T\cos\theta_0\sin\phi_0 \xi^2+(1+r_0 \cos\theta_0)[2\xi \cos\phi_0 \sin^2(T\sqrt{\xi})+(r_0+\cos\theta_0)\sin\phi_0(2T\xi-\sqrt{\xi}\sin(2T\sqrt{\xi})) ],\nonumber\\
			&&\sin\theta_0 [2r_0^2T(2+r_0^2+\cos(2\theta_0))+\sqrt{\xi}\sin(2T\sqrt{\xi})+r_0 \cos\theta_0(2T+6r_0^2T+\sqrt{\xi}\sin(2T\sqrt{\xi}))]\Big\}\nonumber\\
			&=&\{\cos \phi_0, \sin \phi_0,0\}\;,
		\end{eqnarray}
		where $\xi:=1+r_0^2+2r_0 \cos\theta_0,\mathcal{E}:=2\xi\sqrt{r_0^2T^2\xi \sin^2\theta_0+(1+r_0 \cos\theta_0)^2\sin^2(T\sqrt{\xi})}$.
For the SSH model of Eq.~(\ref{eq:SSH}), $\vec{e}_k$ associated with the initial values $\{v_0, w_0, k_0\}$ around the TPT point can be simplified into
		\begin{eqnarray}
			\lim\limits_{\substack{k \to \pi \\ w \to v}}\vec{e}_k&=&\frac{1}{\mathcal{X}}\Big\{w_0\sin k_0[2T^2v_0^2(v_0^2+w_0^2(2+\cos(2k_0)))+w_0^2\sqrt{\chi}\sin(2T\sqrt{\chi})+v_0 w_0 \cos(k_0)(2T(3v_0^2+w_0^2)+\sqrt{\chi}\sin(2T\sqrt{\chi}))],\nonumber\\
			&&-2Tw_0\cos(k_0) \chi^2+\sqrt{\chi}w_0(w_0+v_0 \cos k_0)(v_0+w_0 \cos k_0)(2T\sqrt{\chi}-\sin(2T\sqrt{\chi})),\nonumber\\
			&&2\chi w_0 (w_0+v_0 \cos k_0)\sin^2(T\sqrt{\chi})\Big\}\nonumber\\
			&=&\{0,1,0\}\;,
		\end{eqnarray}
		where $\chi:=v_0^2+w_0^2+2v_0w_0\cos k_0$, $\mathcal{X}:=2\chi\sqrt{w_0^2[T^2v_0^2\chi\sin^2 k_0+(w_0+v_0 \cos k_0)^2 \sin^2(T \sqrt{\chi})]}$.

\section{Numerical simulations for adaptive multiparameter estimation strategy}\label{App:sim}
For the adaptive multiparameter estimation scheme described in Sec.~\ref{subsec:model1}, a set of numerical simulations are provided in Figs.~\ref{Figmodel1}(a) and (b) to assess the impacts of experimental defects on the estimate precision of $\theta$ and $r$. The relevant data are presented in Tables~\ref{Table}.
\begin{figure}[!h]
	\centering
	\includegraphics[width=0.65\textwidth]{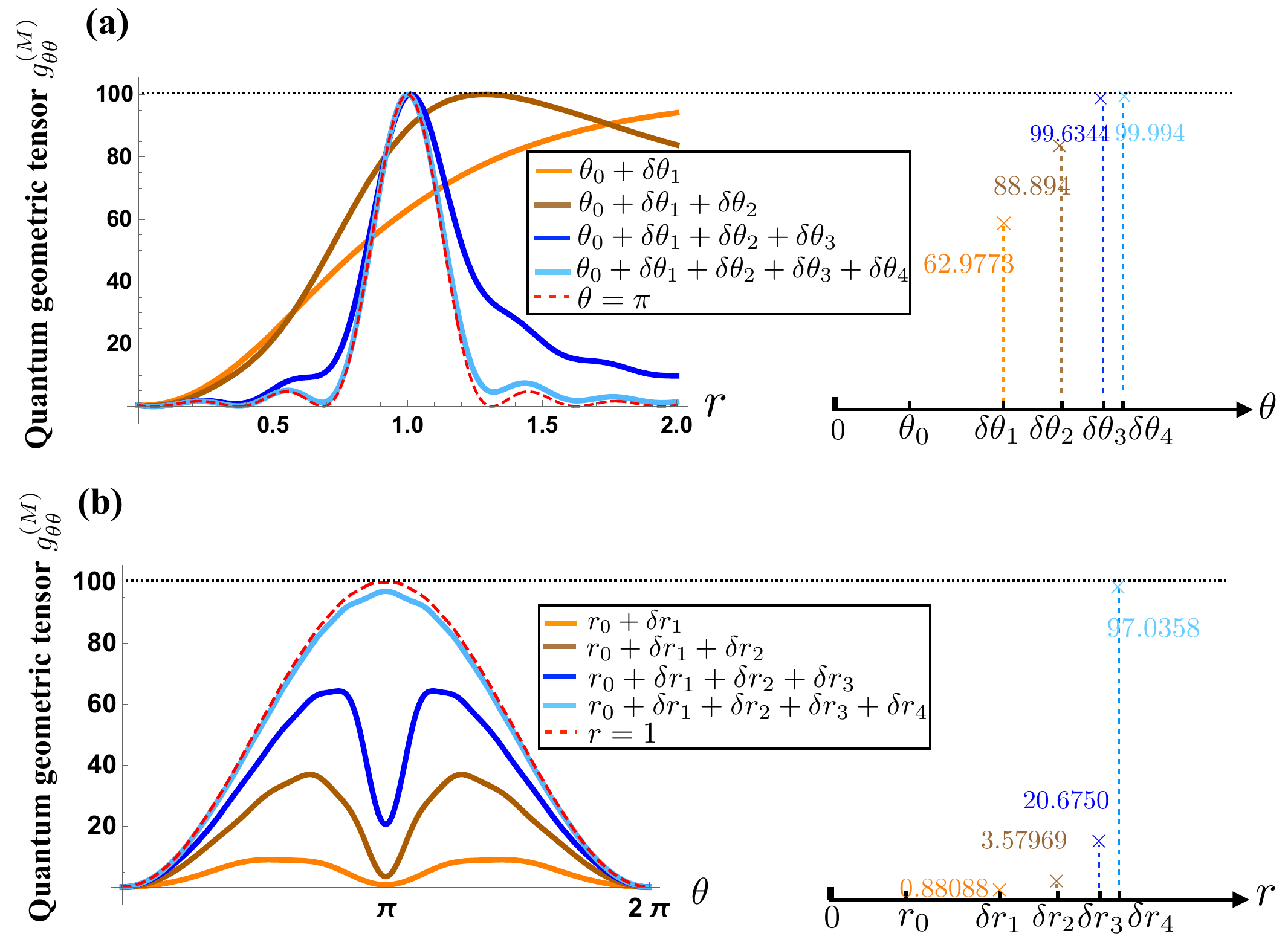}
	\caption
	{
	Results of the QMT $g_{\theta \theta}^{(M)}$ obtained from an adaptive mutiparameter estimation scheme (see  Sec.~\ref{subsec:model1}) are displayed in panel (a) (panel (b)) with the given value of $N_\theta=4$ ($N_r=4$), respectively. 
	In panels (a)-(b), the red dashed curve represents the theoretical QMT result with $\theta=\pi$ ($r=1$), and the other colorful curves individually denote the updated QMT results after the $i (j)$-th time adaptive adjustment for $\theta$ ($r$) for $i, j=\{1,2,3,4\}$, respectively.
	The desired QCRB of $\theta$ (i.e. $g_{\theta \theta}^{(M)}=T^2=100$) is marked by the black dashed line.
	Here $T=10$ is set for the simulation.
	}
	\label{Figmodel1}
\end{figure}

\begin{table*}[h]
	\centering
	\caption{\label{table:with}Simulation data for Fig.~\ref{Figmodel1}(a) and (b)}
	\begin{ruledtabular}
    \renewcommand{\arraystretch}{1.2}
		\begin{tabular}{c|c|c|c|c}
			\textbf{Initial value}&\textbf{Accumulated step-length}&\textbf{Estimated value}&\textbf{Deviation from} &\textbf{QMT at}\\
			&&&\textbf{the initial value} &\textbf{the TPT point}\\
			\hline
			&$\delta \theta_1=\frac{\pi}{3}$&$\theta_0 \simeq \pi-\frac{\pi}{3}$&1.309&62.9773\\
			$\theta_0=\frac{\pi}{4}$&$\delta \theta_1=\frac{\pi}{3},\delta \theta_2=\frac{\pi}{5}$&$\theta_0 \simeq \pi-\left(\frac{\pi}{3}+\frac{\pi}{5}\right)$&0.680&88.894\\
			&$\delta \theta_1=\frac{\pi}{3},\delta \theta_2=\frac{\pi}{5},\delta\theta_3=\frac{\pi}{6}$&$\theta_0 \simeq \pi-\left(\frac{\pi}{3}+\frac{\pi}{5}+\frac{\pi}{6}\right)$&0.157&99.6344\\
			&$\delta \theta_1=\frac{\pi}{3},\delta \theta_2=\frac{\pi}{5},\delta\theta_3=\frac{\pi}{6},\delta \theta_4=\frac{\pi}{15}$&$\theta_0 \simeq \pi-\left(\frac{\pi}{3}+\frac{\pi}{5}+\frac{\pi}{6}+\frac{\pi}{15}\right)$&0.052&99.994\\
			\hline
			&$\delta r_1=0.1$&$r_0 \simeq 1-0.1$&0.7&0.88088\\
			$r_0=0.2$&$\delta r_1=0.1,\delta r_2=0.3$&$r_0 \simeq 1-(0.1+0.3)$&0.4&3.57969\\
			&$\delta r_1=0.1,\delta r_2=0.3,\delta r_3=0.2$&$r_0 \simeq 1-(0.1+0.3+0.2)$&0.2&20.6750\\
			&$\delta r_1=0.1,\delta r_2=0.3,\delta r_3=0.2,\delta r_3=0.17$&$r_0 \simeq 1-(0.1+0.3+0.2+0.17)$&0.03&97.0358\\
		\end{tabular}
	\end{ruledtabular}
\label{Table}
\end{table*}
\end{widetext}


\begin{thebibliography}{100}
\bibitem{Changhao2022}
C. Li, M. Chen, and P. Cappellaro, {\it A geometric perspective: experimental evaluation of the quantum Cramer-Rao bound}, arXiv:2204.13777.
\bibitem{Cai2022}
M. Yu, X. Li, Y. Chu, B. Mera, F. N. Ünal, P. Yang, Y.  Liu, N. Goldman and J. Cai, {\it Experimental demonstration of topological bounds in quantum metrology}, arXiv:2206.00546.
\bibitem{Review_Goldman_2022}
B. Mera, A. Zhang, and N. Goldman, {\it Relating the topology of Dirac Hamiltonians to quantum geometry: When the quantum metric dictates Chern numbers and winding numbers}, SciPost Phys. {\bf 12}, 018 (2022).
\bibitem{Provost1980}
J. P. Provost, and G. Vallee, {\it Riemannian structure on manifolds of quantum states}, Commun.  Math.  Phys. {\bf 76}, 289 (1980).
\bibitem{ReviewCarollo}
A. Carollo, D. Valenti, and B. Spagnolo, {\it Geometry of quantum phase transitions}, Phys. Rep. {\bf 838}, 1 (2020).
\bibitem{Zeng2023}
B. Xia, J. Huang, H. Li, H. Wang, and G. Zeng, {\it Toward incompatible quantum limits on multiparameter estimation}, Nature Commun. {\bf14}, 1021 (2023).
\bibitem{critical}
G. D. Fresco, et al., {\it Multiparameter quantum critical metrology}, SciPost Phys., {\bf 13}, 077 (2022).
\bibitem{critical_review}
R. D. Candia, et al., {\it Critical parametric quantum sensing}, npj Quantum Inform., {\bf  9}, 23 (2023).
\bibitem{critical_new}
A. McDonald, and A. A. Clerk, {\it Exponentially-enhanced quantum sensing with non-Hermitian lattice dynamics}, Nature Commun. {\bf 11}, 5382 (2020).
\bibitem{Ozawa2021}
T. Ozawa, and B. Mera, {\it Relations between topology and the quantum metric for Chern insulators}, Phys. Rev. B  {\bf 104}, 045103 (2021).
\bibitem{Kolodrubetz2017}
M. Kolodrubetz, D. Sels, P. Mehta, and A. Polkovnikov, {\it Geometry and non-adiabatic response in quantum and classical systems}, Phys. Rep. {\bf 697}, 1 (2017).
\bibitem{Demirplak2005}
M. Demirplak and S. A. Rice, {\it Assisted adiabatic passage revisited}, J.  Phys. Chem. B  {\bf 109}, 6838 (2005).
\bibitem{Cai2019} 
M. Yu, et al., {\it Experimental measurement of the quantum geometric tensor using coupled qubits in diamond}, Natl. Sci. Rev. {\bf 7}, 254 (2019).
\bibitem{Yu2019}
X.  Tan, et al., {\it Experimental Measurement of the Quantum Metric Tensor and Related Topological Phase Transition with a Superconducting Qubit}, Phys. Rev. Lett. {\bf 122}, 210401 (2019).
\bibitem{Klees2020}
R. L. Klees, G. Rastelli,  J. C. Cuevas, and W. Belzig, {\it Microwave Spectroscopy Reveals the Quantum Geometric Tensor of Topological Josephson Matter}, Phys. Rev. Lett.  {\bf 124}, 197002 (2020).
\bibitem{Flaschner2016}
N. Fläschner, B. S. Rem, M. Tarnowski, D. Vogel, D.-S. Lühmann, K. Sengstock, and C. Weitenberg, {\it Experimental reconstruction of the Berry curvature in a Floquet Bloch band}, Science {\bf 352}, 1091 (2016).
\bibitem{Matsumoto2002}
K. Matsumoto,  {\it A new approach to the Cramér-Rao-type bound of the pure-state model}, J. Phys. A: Math. Gen. {\bf 35}, 3111 (2002).
\bibitem{Review2019}
J. Liu, H. Yuan, X.-M. Lu, and X. Wang, {\it Quantum Fisher information matrix and multiparameter estimation}, J. Phys. A: Math. Theor. {\bf53}, 023001 (2019).
\bibitem{Albarelli2020}
F. Albarelli, M. Barbieri, M. G. Genoni, and I. Gianani, {\it A perspective on multiparameter quantum metrology: From theoretical tools to applications in quantum imaging}, Phys. Lett. A {\bf 384}, 126311 (2020).
\bibitem{Kok2021}
J. S. Sidhu, Y. Ouyang, E. T. Campbell, and P. Kok, {\it Tight Bounds on the Simultaneous Estimation of Incompatible Parameters, } Phys. Rev. X {\bf 11}, 011028 (2021).
\bibitem{Albarelli2022}
F. Albarelli,  and R. Demkowicz-Dobrza\ifmmode $\acute${n}\else \'{n}\fi{}ski, {\it Probe Incompatibility in Multiparameter Noisy Quantum Metrology}, Phys. Rev. X {\bf 12}, 011039 (2022).
\bibitem{Belliardo2021}
F. Belliardo, and V. Giovannetti, {\it Incompatibility in quantum parameter estimation}. New J.  Phys. {\bf 23}, 063055 (2021).
\bibitem{Chen2022}
H. Chen, Y. Chen, and H. Yuan, {\it Incompatibility measures in multiparameter quantum estimation under hierarchical quantum measurements}. Phys. Rev. A {\bf 105}, 062442 (2022).     
\bibitem{Wang2016}
W. Guo, W. Zhong, X.-X. Jing, L.-B. Fu, and X. Wang, {\it Berry curvature as a lower bound for multiparameter estimation}, Phys. Rev. A  {\bf 93}, 042115 (2016).
\bibitem{Lu2021}
X.-M. Lu,  and X. Wang, {\it Incorporating Heisenberg's Uncertainty Principle into Quantum Multiparameter Estimation}, Phys. Rev. Lett. {\bf 126}, 120503 (2021).
\bibitem{Yuan2016}
H. Yuan, {\it Sequential Feedback Scheme Outperforms the Parallel Scheme for Hamiltonian Parameter Estimation},  Phys. Rev. Lett.  {\bf 117}, 160801 (2016).
\bibitem{Hou2021}
Z. Hou, J.-F. Tang, H. Chen, H. Yuan, G.-Y. Xiang, C.-F. Li, and G.-C. Guo, {\it Zero–trade-off multiparameter quantum estimation via simultaneously saturating multiple Heisenberg uncertainty relations}, Sci. Adv. {\bf 7}, eabd2986 (2021).
\bibitem{Hou2021PRL}
Z. Hou, Y. Jin, H. Chen,  J.-F. Tang, C.-J. Huang, H. Yuan, G.-Y. Xiang, C.-F. Li, and G.-C. Guo,{\it  ``Super-Heisenberg'' and Heisenberg Scalings Achieved Simultaneously in the Estimation of a Rotating Field}, Phys. Rev. Lett. {\bf 126}, 070503 (2021).
\bibitem{Yang2022}
Y. Yang, S. Ru, M. An, Y. Wang, F. Wang, P. Zhang, and F. Li, {\it Multiparameter simultaneous optimal estimation with an SU(2) coding unitary evolution}, Phys. Rev. A {\bf 105}, 022406 (2022).
\bibitem{Note2022}
Here we consider the general $\lambda$-dependent quantum state manifold and does not restrict in the ground state manifold.
\bibitem{Carollo2019}
A. Carollo, B. Spagnolo, A. A. Dubkov, and D. Valenti, {\it On quantumness in multi-parameter quantum estimation}, J.  Stat. Mech. Theor. E.  {\bf 2019}, 094010 (2019).
\bibitem{momentum}
Y.-Q. Ma, S.-J. Gu, S. Chen, H. Fan, and W.-M. Liu, {\it The Euler number of Bloch states manifold and the quantum phases in gapped fermionic systems}, Europhys. Lett. {\bf 103}, 10008 (2013).
\bibitem{Pang2014}
S. Pang, and T. A. Brun, {\it Quantum metrology for a general Hamiltonian parameter}, Phys. Rev. A {\bf 90}, 022117 (2014).
\bibitem{coarse}
Y. Yang. L. Xu, and V. Giovannetti, {\it Two-parameter Hong-ou-Mandel dip}, Sci. Reports {\bf 9}, 10821 (2019).
\bibitem{coarsenew}
Y. Yang. L. Xu, and V. Giovannetti, {\it Exclusive Hong-Ou-Mandel zero-coincidence point}, Phys. Rev. A {\bf 100}, 063810 (2019).
\bibitem{Goldman2018}
G. Palumbo, and N. Goldman, {\it Revealing Tensor Monopoles through Quantum-Metric Measurements}, Phys. Rev. Lett. {\bf 121}, 170401 (2018).
\bibitem{topological_book}
J. K. Asbóth, L.  Oroszlány, and A.  Pályi, {\it A short course on topological insulators: Band Structure and Edge States in One and Two Dimensions}, Springer International Publishing (2016).
\bibitem{Ozawa2018}
T. Ozawa, and N. Goldman, {\it Extracting the quantum metric tensor through periodic driving}, Phys. Rev. B {\bf 97}, 201117(R) (2018).
\bibitem{Adaptive2021}
H. Xu, L.  Wang, H.  Yuan, and X. Wang, {\it Generalizable control for multiparameter quantum metrology}, Phys. Rev. A {\bf 103}, 042615 (2021).
\bibitem{Chu2023}
Y. Chu, Y., X. Li, and J. Cai, {\it Strong Quantum Metrological Limit from Many-Body Physics}. Phys. Rev. Lett. {\bf 130}, 170801 (2023).
\bibitem{Ying2022}
Z.-J. Ying, S. Felicetti, G. Liu, and D. Braak, {\it Critical Quantum Metrology in the Non-Linear Quantum Rabi Model}, Entropy  {\bf 24}, 1015 (2022).
\bibitem{Garbe2020}
L. Garbe, M. Bina, A. Keller, M. G. A Paris, and S. Felicetti, {\it Critical Quantum Metrology with a Finite-Component Quantum Phase Transition}, Phys. Rev. Lett. {\bf 124}, 120504 (2020).
\bibitem{WeipingZhang2023}
L. Zhou, J. Kong, Z. Lan, and W. Zhang, {\it Dynamical quantum phase transitions in a spinor Bose-Einstein condensate and criticality enhanced quantum sensing}, Phys. Rev. Res. {\bf 5}, 013087 (2023).
\bibitem{Yin2019}
S. Yin, J. Song, Y. Zhang, and S. Liu, {\it Quantum Fisher information in quantum critical systems with topological characterization}, Phys. Rev. B {\bf 100}, 184417 (2019).
\end{thebibliography}
\end{document}